\documentclass[reprint, amsmath, amssymb, aps]{revtex4-1}

\usepackage{graphicx}
\usepackage{dcolumn}
\usepackage{bm}
\usepackage{placeins}
\usepackage{float} 

\begin{document}

\preprint{PHYSICAL REVIEW E}

\title{Effects of time-periodic forcing in a Cahn-Hilliard model for Langmuir-Blodgett transfer}
\author{Phong-Minh Timmy Ly$^{1}$}
\email{timmy.ly@uni-muenster.de}
\author{Uwe Thiele$^{1,2}$}
\author{Lifeng Chi$^{3}$}
\author{Svetlana V. Gurevich$^{1,2}$}
\affiliation{$^{1}$Institute for Theoretical Physics, University of M\"unster, Wilhelm-Klemm-Str. 9, D-48149 M\"unster, Germany}
\affiliation{$^{2}$Center for Nonlinear Science (CeNoS), University of M\"unster, Corrensstrasse 2, D-48149 M\"unster, Germany}
\affiliation{$^{3}$Institute of Functional Nano \& Soft Materials (FUNSOM) and Jiangsu Key Laboratory for Carbon-based Functional Materials \& Devices Collaborative Innovation Center of Suzhou Nano Science and Technology, Soochow University, 215123 Suzhou, P. R. China}

\begin{abstract}

\begin{description}
	\item[Abstract]

	The influence of a temporal forcing on the pattern formation in Langmuir-Blodgett transfer is studied employing a generalized Cahn-Hilliard model. The occurring frequency locking effects allow for controlling the pattern formation process. In the case of one-dimensional (i.e., stripe) patterns one finds various synchronization phenomena such as entrainment between the distance of deposited stripes and the forcing frequency. In two dimensions, the temporal forcing gives rise to the formation of intricate complex patterns such as vertical stripes, oblique stripes and lattice structures. Remarkably, it is possible to influence the system in the spatial direction perpendicular to the  forcing direction leading to synchronization in two spatial dimensions. 
\end{description}

\end{abstract}

\pacs{Valid PACS appear here}
\maketitle

\section{\label{sec:intro}Introduction}
An adjustment of rhythms of oscillating objects due to their weak interaction, i.e., synchronization, is often encountered in everyday life. Examples are the adjustment of the clapping of hands in a crowd or of the chirping of male crickets to attract females \cite{Walker891}.
This nonlinear phenomenon of synchronization is investigated in many research fields, for systems ranging from biological oscillators \cite{MirolloStrogatz:1990,glass2001synchronization} to technical applications like power grids \cite{FilatrellaNielsenPedersen:2008}, chaotic pulsed lasers~\cite{SugawaraPRL1994}, signal encryption \cite{CuomoOppenheimStrogatz:1993,KeuninckxSorianoFischerRMirassoNguimdoVanderSande:2017} and nano-oscillators~\cite{DemidovNature2014}. Indeed, a periodically applied force can cause a simple nonlinear oscillator to become entrained at a frequency that is rationally related to the applied frequency, a phenomenon known as \emph{frequency-locking}~\cite{PikovskyRosenblumKurths:2003}. Spatially extended systems can exhibit a similar frequency locking behavior. The presence of time-periodic forcing does not only allow one to create new kinds of patterns like, e.g, standing waves induced by the Faraday instability in vertically shaken fluid layers~\cite{Faraday1931} or oscillons in vertically shaken layers of granular media~\cite{UmbanhowarNature1996}, but also to control the patterns by varying the forcing parameters, see e.g.~\cite{Mikhailov2006}. We mention periodically driven spatially extended chemical systems~\cite{PetrovOuyangSwinney:1997,LinPRL2000}, the forced complex Ginzburg-Landau equation~\cite{CHATE199917,RudzickPRL2006,YOCHELIS2004201,ParkPRL2001,ElphickPRL98}, periodically modulated Bose condensates~\cite{EngelsPRL2007} and mode-locking in fibre optical systems due to the modulation of a dissipative parameter~\cite{TarasovNature2016}. 

The spatial counterpart of frequency locking, \emph{wavenumber locking}, has also been of increasing interest in recent years. Much work is devoted to pattern-forming systems that are subjected to a \emph{spatially periodic forcing} like spatially periodic illumination of a chemical reaction~\cite{dolnik2011locking}, electroconvection of a nematic liquid crystal in a spatially periodic electric field~\cite{lowe1983commensurate} and Rayleigh-B\'enard convection on a plate with a parallel-ridge topography~\cite{mccoy2008self}. Extensive theoretical studies are performed in the context of a Ginzburg-Landau type equation~\cite{coullet1986commensurate}, Swift-Hohenberg equations~\cite{manor2008wave}, dewetting and phase separation phenomena~\cite{TBBB2003epje,krekhov2004phase,krekhov2009formation}, and dip-coating systems \cite{KoepfGurevichFriedrich:2011,WilczekGurevich:2014,ZhuWilczekHirtz:2016,WilczekZhuChiThieleGurevich:2017}. The combination of spatial and temporal forcing is also considered~\cite{RuedigerPRL2003,UtznyEPL2002}, see also~\cite{RUDIGER200773} for a review. Note that in general, spatially forced extended systems can respond in two or three spatial dimensions, even if the spatial forcing is only one-dimensional, while locking in the time domain is inherently one-dimensional. 

\begin{figure}[tb!]
	\includegraphics[width = 0.45\textwidth]{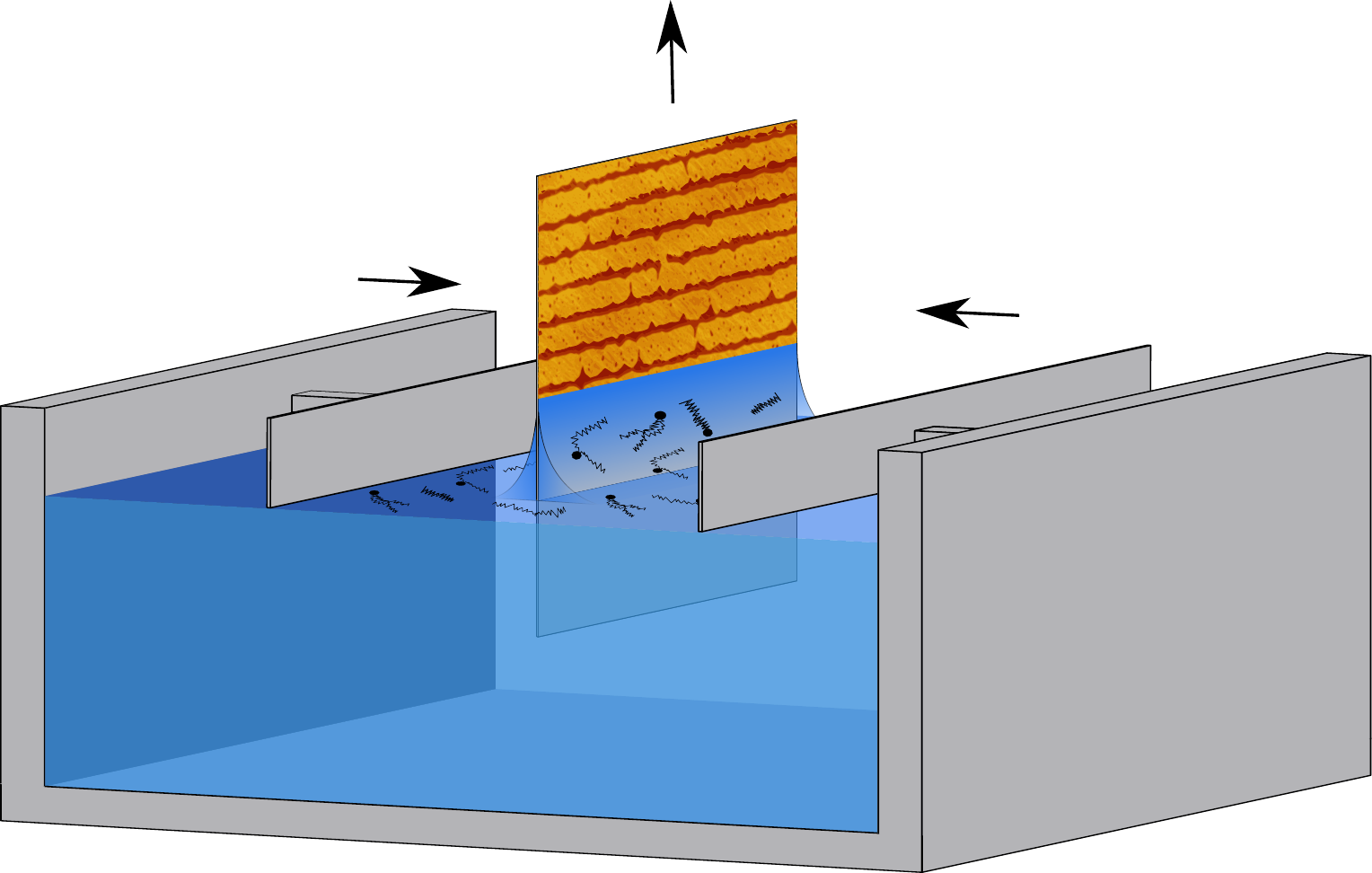}
	\caption{\label{fig:lb_setup} Sketch of the experimental setup of the Langmuir-Blodgett (LB) transfer. A solid substrate is pulled out of a liquid bath covered by a monolayer of surfactants at constant density. The latter is ensured by the use of movable barriers limiting the bath surface. On the liquid surface the surfactant is in the liquid-expanded (LE) phase. The layer may undergo substrate-mediated condensation in the vicinity of the three-phase contact line and is then deposited onto the substrate (partly) in the liquid-condensed (LC) phase.}
\end{figure}

Here, we are interested in the effects of temporal forcing on self-organized deposition patterns occurring in Langmuir-Blodgett (LB) transfer~\cite{ChenLenhertHirtzLuFuchsChi:2007}. We employ a generalized Cahn-Hilliard-type model \cite{KoepfGurevichFriedrichThiele:2012,WilczekGurevich:2014,KoepfThiele:2014} to investigate the resulting locking effects. In particular, we show how this allows us to use time-periodic forcing to control deposition patterns. The experimental setup is sketched in Fig.~\ref{fig:lb_setup}. It consists of a liquid bath, a so-called LB trough \cite{Blodgett:1935}, that is covered by a floating monolayer of surfactants, i.e., amphiphilic molecules, specifically: dipalmitoylphosphatidylcholine (DPPC). When pulling a solid substrate out of the bath, a thin liquid film covered by surfactants is deposited on the substrate. Concurrently, movable barriers at the bath surface are moved along it in order to keep the surface pressure constant (and therefore the surfactant density and phase). Once the liquid layer is deposited the solvent evaporates and a nearly dry, possibly structured, layer of surfactants covers the solid substrate. The region where the bath surface approaches the substrate corresponds to a three-phase contact region and physico-chemical mechanisms related to wettability may influence the transfer.

In the considered experiments, the surfactant on the bath (i.e., before the transfer) is kept in the so-called liquid-expanded phase \cite{kaganer:1999} where the molecules are spread out on the liquid surface in a disordered manner. 
However, in the vicinity of the three-phase contact line, changes in the phase behavior of the surfactant monolayer can be induced by interactions with the nearby substrate. It is found that in this way a first order phase transition from the liquid-expanded (LE) to the liquid-condensed (LC) phase can occur. This is termed substrate-mediated condensation (SMC) \cite{RieglerSpratte:1992, SpratteRiegler:1994, SpratteChiRiegler:1994, GrafRiegler:1998, LeporattiBrezesinskiMoehwald:2000}. In the LC phase the elongated surfactant molecules are densely ordered and nearly perpendicular to the substrate with their hydrophilic heads pointing towards it, whereas molecules of the LE phase are spread out and disordered. Depending on the transfer velocity, a spatially homogeneous or inhomogeneous (patterned) surfactant layer is transferred onto the moving substrate. At low and high velocities a homogeneous layer of LE and LC phase is transfered, respectively, whereas at intermediate velocities periodic patterns of the two phases occur. In the simplest case, this results in patterns of stripes arranged perpendicular to the transfer direction (horizontal stripes, cf.~Fig.~\ref{fig:lb_setup}). Further, there exists a small region at low transfer velocities where the horizontal stripes are transversally unstable and stripes emerge that are parallel to the transfer direction (vertical stripes). 
In recent years, the system has been extensively investigated. Experiments achieve patterns with typical lengths which range from nano- to microscales that are homogeneous over square centimeter surface areas. \cite{Gleiche2000Nanoscopic,ChenLenhertHirtzLuFuchsChi:2007,LiKoepfGurevichFriedrichChi:2012}. On the theoretical side, well performing models for the LB system have been developed. There is a hydrodynamic thin-film model describing the coupled time evolution of the surfactant density and film height \cite{KoepfGurevichFriedrich:2009,KoepfGurevichFriedrichChi:2010,KoepfGurevichFriedrichThiele:2012}. This model can be brought into a gradient dynamics form \cite{ThAP2012pf,ThAP2016prf,Thie2018csa} allowing for systematic extensions based on improvements of the underlying energy functional. The analysis of the hydrodynamic model shows that main features do only weakly depend on details of the film height profile allowing one to develop a simpler generic model. The resulting generalized Cahn-Hilliard-type model captures the main features of SMC-induced patterning in LB transfer \cite{KoepfGurevichFriedrichThiele:2012,KoepfThiele:2014,WilczekTewesGurevichKoepfChiThiele:2015}. 
The simpler model then allows for extensive scans of parameter space via time simulations \cite{KoepfGurevichFriedrichThiele:2012,WilczekTewesGurevichKoepfChiThiele:2015} and the use of numerical continuation techniques to analyse the bifurcation structure of the problem \cite{KoepfGurevichFriedrichThiele:2012,KoepfThiele:2014}. Mathematical discussions of underlying generic patterning fronts in Cahn-Hilliard-type systems can be found in \cite{GoSc2015arma,GoSc2016n}.

With the LB technique one can produce extended surfaces patterned with small-scale structures. The technique is cheaper than classically used lithographic methods since the patterns are formed via self-organization and neither masks nor writing devices like lasers are required. However, the patterns created by classical LB transfer are prone to defects like the growth of side-branches which results in non-uniform patterns. An effective method of suppressing such unwanted structures is the use of spatial forcing in the form of prestructured substrates, e.g., substrates with gold stripes which periodically modulate its wetting properties \cite{ZhuWilczekHirtz:2016}. Then, in LB transfer various locking effects occur where the periodicity of the gold stripes stands in a rational ratio to the periodicity of the deposited stripe pattern of surfactant phases. This provides a further means of control of the patterns~\cite{KoepfGurevichFriedrich:2011,WilczekGurevich:2014,ZhuWilczekHirtz:2016,WilczekZhuChiThieleGurevich:2017}. 

However, although the use of prestructured substrates is a powerful tool of patterning control in LB transfer, it also has disadvantages in real-world applications. The most serious one is the effort that is needed to fabricate the prestructured substrates: First, a poly(methylmethacrylate) (PMMA) layer is spin-coated onto the silicon substrate as a resist. The form of the prestructure is then written into the PMMA layer by electron beam lithography. The substrate is then developed, thereby removing the surplus PMMA. Afterwards, a 3~nm thick chrome adhesion layer and the actual 20~nm thick gold prestructure are deposited by thermal evaporation under vacuum conditions~\cite{ZhuWilczekHirtz:2016,WilczekZhuChiThieleGurevich:2017}. As a result, the effort of producing the prestructure easily exceeds the one of the actual LB transfer process. Therefore, it is highly desirable to develop alternative control techniques that are simpler and equally powerful.

Here, we explore an alternative option for a control technique based on temporal modulations of control parameters of the transfer process. First steps in this direction were performed in~\cite{Wilczek:2016}. Possibly the most important control parameter is the transfer velocity, i.e., the velocity at which the substrate is withdrawn from the liquid bath. It influences both, the type of deposited pattern (horizontal vs.\ vertical stripes) and its properties (e.g., period). In particular, we show that a temporally periodic modulation of the transfer velocity can lead to very similar locking effects as observed for prestructured substrates. Our work is structured as follows: In Section~\ref{sec:model} the Cahn-Hilliard type model is introduced, followed by the presentation of results of numerical simulations for one-dimensional substrates in Section~\ref{sec:1d} and for two-dimensional substrates in Section~\ref{sec:2d}. Finally, a summary of the main findings and an outlook are presented in Section~\ref{sec:conclusion}.

\section{Theoretical model}\label{sec:model}
As explained in Sec.~\ref{sec:intro}, the formation of domains of the LC phase only occurs in the close vicinity of the substrate. Therefore, the most relevant part of the experiment that needs to be described by a theoretical model corresponds to the surrounding of the three-phase contact line, where the transition from the deep water bath to the thin deposited surfactant-covered layer occurs. Under this assumption, the transfer process can be well described by the dynamics of the water layer in a long-wave (or lubrication) approximation~\cite{OronDavisBankoff:1997,SADF2007jfm,GTLT2014prl} coupled to the dynamics of the surfactant at its surface \cite{CrMa2009rmp,ThAP2012pf}. A specific hydrodynamic long-wave model that accounts for the phase transition in the surfactant monolayer is developed in ref.~\cite{KoepfGurevichFriedrich:2009} and describes most phenomena occurring during LB transfer. However, it turns out that the meniscus shape in the vicinity of the contact line is rather static when a homogeneous layer of surfactants is deposited and only shows small oscillations for transfer velocities within the regime of patterned deposition. In fact, these meniscus oscillations do not significantly influence the mechanism of pattern formation as confirmed by numerically 'freezing' the meniscus \cite{KoepfGurevichFriedrichThiele:2012}. Therefore, it is possible to eliminate the description of the film height dynamics from the model and directly incorporate its effect on the substrate-mediated condensation (SMC) into the description of the surfactant dynamics in the form of a position-dependent chemical potential. As a result a generalized Cahn-Hilliard-type model \cite{CahnHilliard:1958,Cahn1965jcp}  is developed \cite{KoepfGurevichFriedrichThiele:2012} that solely describes the dynamics of an order parameter  $c(\mathbf{x},t)$ that corresponds to a scaled surfactant density. The nondimensionalized model reads 
\begin{equation}
	\partial_t c(\mathbf{x}, t) = \nabla\cdot\left(\nabla\frac{\delta F[c]}{\delta c}\right)
	\label{eq:ch}
\end{equation}
with the free energy functional
\begin{equation}
	F[c] = \int \frac{1}{2}(\nabla c)^2 + f(c,x)\;d\mathbf{x}\,,
	\label{eq:free_energy}
\end{equation}
where
\begin{equation}
	f(c,x) = - \frac{1}{2}c^2 + \frac{1}{4}c^4 + \mu\zeta(x-x_s)\,c\;.
	\label{eq:local_free_energy}
\end{equation}
Here, $F[c]$ includes the classical Cahn-Hilliard contributions \cite{CahnHilliard:1958}, namely, a square-gradient term that penalizes spatial inhomogeneities, and a double-well approximation for the free energy of the uniform state, which is justified in the vicinity of the LE-LC phase transition of the monolayer. The minima represent the LE ($c = -1$) and LC ($c = 1$) phase. The onset of SMC at the three-phase contact line is modeled by a spatially varying external field $\mu\zeta(x-x_s)$ where $\mu$ is a prefactor setting the strength of the SMC-field and $x_s$ represents the contact line position. 
In the bath region ($x < x_s$) the two phases are energetically equivalent, i.e., the interaction with the substrate tends to zero for: $\zeta(x\to-\infty) \to 0$, whereas for the deposited layer ($x > x_s$) one has $\zeta(x\to\infty) \to -1$. This implies that there the double-well potential is tilted and the LC phase is energetically prefered, while the LE phase is disfavored. The two regions of different behaviour are smoothly connected by a hyperbolic tangent with a transition length of $l_s$ centered about $x_s$
\begin{equation}
	\zeta(x-x_s) = -\frac{1}{2}\left[1 + \tanh\left(\frac{x-x_s}{l_s}\right)\right]. 
	\label{eq:zeta}
\end{equation}
The analytical expression used for $\zeta(x-x_s)$ can have any form that monotonically interpolates between the two phases. In particular, we do expect that the specific value of $l_s$ has nearly no influence on the results presented in Sec.~\ref{sec:1d} and Sec.~\ref{sec:2d}. In general, $l_s$ should be chosen neither too small nor too large to avoid the interface becoming too sharp or unphysically wide, respectively.
Inserting  \eqref{eq:free_energy}-\eqref{eq:zeta} into \eqref{eq:ch}, choosing $M = 1$ and adding an advection term, that represents the movement of the substrate with velocity $\mathbf{v}$, results in the generalized Cahn-Hilliard equation 
\begin{equation}
	\partial_t\,c(\mathbf{x},t) = \nabla\cdot\left[\nabla\left(-\Delta c -c + c^3 + \mu\zeta(x-x_s)\right) + 
\mathbf{v} c\right]
	\label{eq:full_ch_eq}
\end{equation}
For all numerical calculations we employ a finite domain $\Omega$ -- in the one-dimensional (1D) case $\Omega_1 = [0,L]$, and in the two-dimensional (2D) case $\Omega_2 = [0,\,L] \times [0,\,L]$. The boundary conditions are
\begin{equation}
	c|_{x=0} = c_0, \quad \partial_x c|_{x=0} = 0, \quad \partial_{xx} c |_{x=0} = 0, 
	\label{eq:bc1}
\end{equation}
\begin{equation}
	\partial_{x} c |_{x=L} = \partial_{xx} c |_{x=L} = 0, \quad c|_{y=0} = c|_{y=L}. 
	\label{eq:bc2}
\end{equation}
They reflect that the surfactant density on the bath is kept constant at a value $c_0$ that represents the LE phase on the liquid bath. For calculations in the 2D case, periodic boundary conditions are used in the $y$-direction. The time-periodic forcing is introduced through the main control parameter of the system which is the transfer velocity \cite{Wilczek:2016}
\begin{equation}
	\mathbf{v} = \bar{v}[1+A\sin(\omega t)] \mathbf{e_x}.
	\label{eq:forcing}
\end{equation}
Here, $\mathbf{v}$ harmonically oscillates about its mean value $\bar{v}$ with an (angular) frequency $\omega$ and forcing amplitude $A$. This will result in frequency entrainment, thereby leading to synchronization effects. In the following, we call Eq.~\eqref{eq:forcing} the forcing term. 

Note that even in the presence of moderate forcing we do not expect qualitative differences in the pattern forming behavior in comparison to the aforementioned full hydrodynamic model. A comparison of a fully hydrodynamic thin-film description and a simplified Cahn-Hilliard model as used here is made in Refs.~\cite{KoepfGurevichFriedrich:2011} and \cite{WilczekGurevich:2014} for the related case of spatially periodic forcing via prestructured substrates. Both model types show similar wavenumber locking phenomena. However, if the prestructure significantly affects the wettability of the substrate and hence the dynamics of the liquid meniscus (i.e., large wettability contrasts are introduced), the contact line position can be affected by the prestructure. In this case, the Cahn-Hilliard model~\eqref{eq:ch} is not applicable any more and the full hydrodynamic model should be used~\cite{WilczekZhuChiThieleGurevich:2017}. 

For the numerical time simulations we employ a NVIDIA CUDA~\cite{NVIDIA:2017} implementation of a second order finite difference scheme for the spatial discretization and a Runge-Kutta 4(5) method with adaptive step size for the time integration~\cite{WilczekGurevich:2014}. To closely model the experimental state at the beginning of the surfactant transfer, if not stated otherwise, the initial conditions are chosen as
\begin{equation}
	c(x,y,t = 0) = c_0 + \frac{1}{2}[1 + \tanh(x-x_s)] (1-c_0) +\;\mathrm{noise}, 
	\label{eq:ic}
\end{equation}
i.e., a simple hyperbolic tangent representing the transition between a homogeneous LE state on the bath and a substrate without surfactant. A small amount of white noise is added.

\section{Solution measures\label{sec:solmeas}}
\begin{figure}[tb]
	\includegraphics[width = 0.49\textwidth]{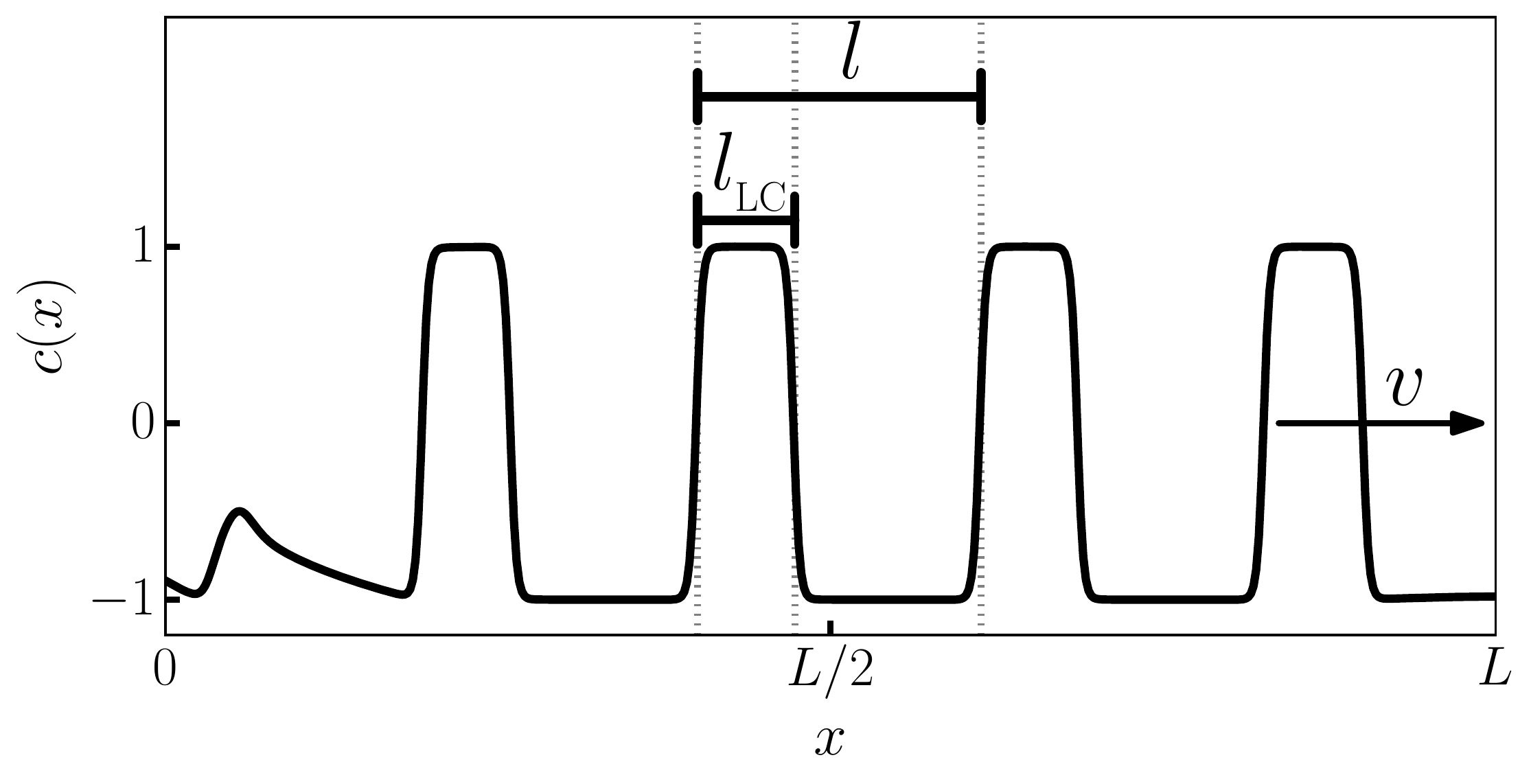}
	\caption{\label{fig:1d_vconst_stripedist} Snapshot of a time-periodic solution of \eqref{eq:full_ch_eq} in the 1D case obtained by direct numerical integration for a domain of size $L$. The stripe period $l$ is defined as the distance between two subsequent rising flanks whereas the width of the stripes of LC phase $l_\mathrm{LC}$ is the distance between the two zero crossings that bracket the region with $c(x)> 0$. The domain size is increased for larger $l$ while maintaining the discretization in order to fully capture the stripes. The black arrow denotes the transfer direction.}
\end{figure}

As our main interest lies in the development of ways to control pattern formation, the following investigation primarily focuses on time-periodic solutions of Eq.~(\ref{eq:full_ch_eq}) that correspond to patterned deposition. In the unforced system, i.e., at constant advection velocity $v=\bar{v}$,  the deposited patterns are mostly horizontal stripes. This implies, that  simulations in the 1D case are sufficient as the resulting solutions are equivalent to cross sections along the transfer direction of solutions in the 2D case. 
To quantify the system behaviour we use three solution measures. The first one is the pattern period $\bar{l}$ defined as the average distance between pairs of subsequent stripes (see Fig.~\ref{fig:1d_vconst_stripedist}). The second one is the average duty cycle $D = \overline{l_\mathrm{LC}/l}$, defined as the average fraction of a period that consists of the LC phase. Here, $l_\mathrm{LC}$  is the width of stripes of the LC phase (Fig.~\ref{fig:1d_vconst_stripedist}). $D$ measures the relative importance of LE and LC phase. 
The third measure is the synchronization order $W$, also known as rotation or winding number. It quantifies the degree of synchronization \cite{PikovskyRosenblumKurths:2003}. In periodically forced systems, $W$ is defined as the ratio between the forcing frequency $\omega$ and the mean frequency $\bar{\Omega}$ of the time-periodic solution exhibited by the system:
\begin{equation}
	W = \frac{\bar{\Omega}}{\omega} = \frac{n}{m} \qquad n,m\in\mathbb{N}. 
	\label{eq:W}
\end{equation}
For strictly periodic dynamics, $W$ is a rational number. It is advantageous to introduce the mean quantities since subsequent pairs of stripes of periodic solutions are usually not identical. Individually, each stripe is characterized by the distance to the next stripe $l_i$, the time intervall to the next stripe $T_i$ (here called ``period''), and frequency $\Omega_i=\frac{2\pi}{T_i}$ (and $l_{\mathrm{LC}_i}$), all with $i\in\mathbb{N}$. This differs from the unforced case where the stripe distance does not vary and the stripe distance directly corresponds to the pattern period. The corresponding quantities we call ``natural''. Synchronization occurs when the mean frequency of the solution $\bar{\Omega}$ is entrained by the forcing frequency $\omega$. In the case of the forced system  \eqref{eq:full_ch_eq}-\eqref{eq:forcing},  individual $\Omega_i$ are determined via the periods $T_i$ which are measured as time intervalls between the passing of two subsequent stripes of a chosen measurement point within the domain, e.g., $x = 0.8L$. On average, the stripes emerge with mean frequency $\bar{\Omega}$, concurrently they advect on average with $\bar{v}$ and it is possible to map the time-periodic solutions onto a circular limit cycle with constant frequency $\bar{\Omega}$. Since after one cycle the solution will on average have advected by one mean stripe distance $\bar{l}$, the radius $r$ of the circle has to be such that $2\pi\cdot r = \bar{l}$. Using $\bar{v} = \bar{\Omega}\cdot r$ for circular motion with constant frequency and eliminating $r$, one arrives at
\begin{equation}
	\bar{l} = \frac{2\pi}{\bar{\Omega}}\bar{v}\;.
\end{equation}
Eliminating $\bar{\Omega}$ by inserting  \eqref{eq:W} leads to
\begin{equation}
	\bar{l} = \frac{2\pi}{\omega W}\bar{v} =: \frac{l_f}{W}\;,
	\label{eq:sync}
\end{equation}
where $l_f$ is the forced stripe distance in the case of $1:1$ synchronization. This relation indicates several ways of controlling the pattern formation. As long as the system is located within a parameter region of constant synchronization order, $\bar{l}$ will be proportional to $\bar{v}$ and inversely related to $\omega$. In consequence, tuning these two parameters allows one to control the mean stripe distance.
One of the factors responsible for the synchronization order is the frequency detuning, which is the difference between the natural frequency $\Omega_{\mathrm{nat}}$ of the system and the forcing frequency $\omega$. The smaller the detuning, the more common are low synchronization orders $W$, i.e., $W= \frac{n}{m}$ resulting from small values of $n$ and $m$. 
Here, $\Omega_{\mathrm{nat}}$ is directly related to the  the natural stripe distance $l_{\mathrm{nat}}=2\pi \bar{v}/\Omega_{\mathrm{nat}}$ that is identical  to the (constant) stripe distance occurring in the unforced system.

\section{Synchronization effects in one dimension}\label{sec:1d}
\begin{figure}[tb!]
	\includegraphics[width = 0.49\textwidth]{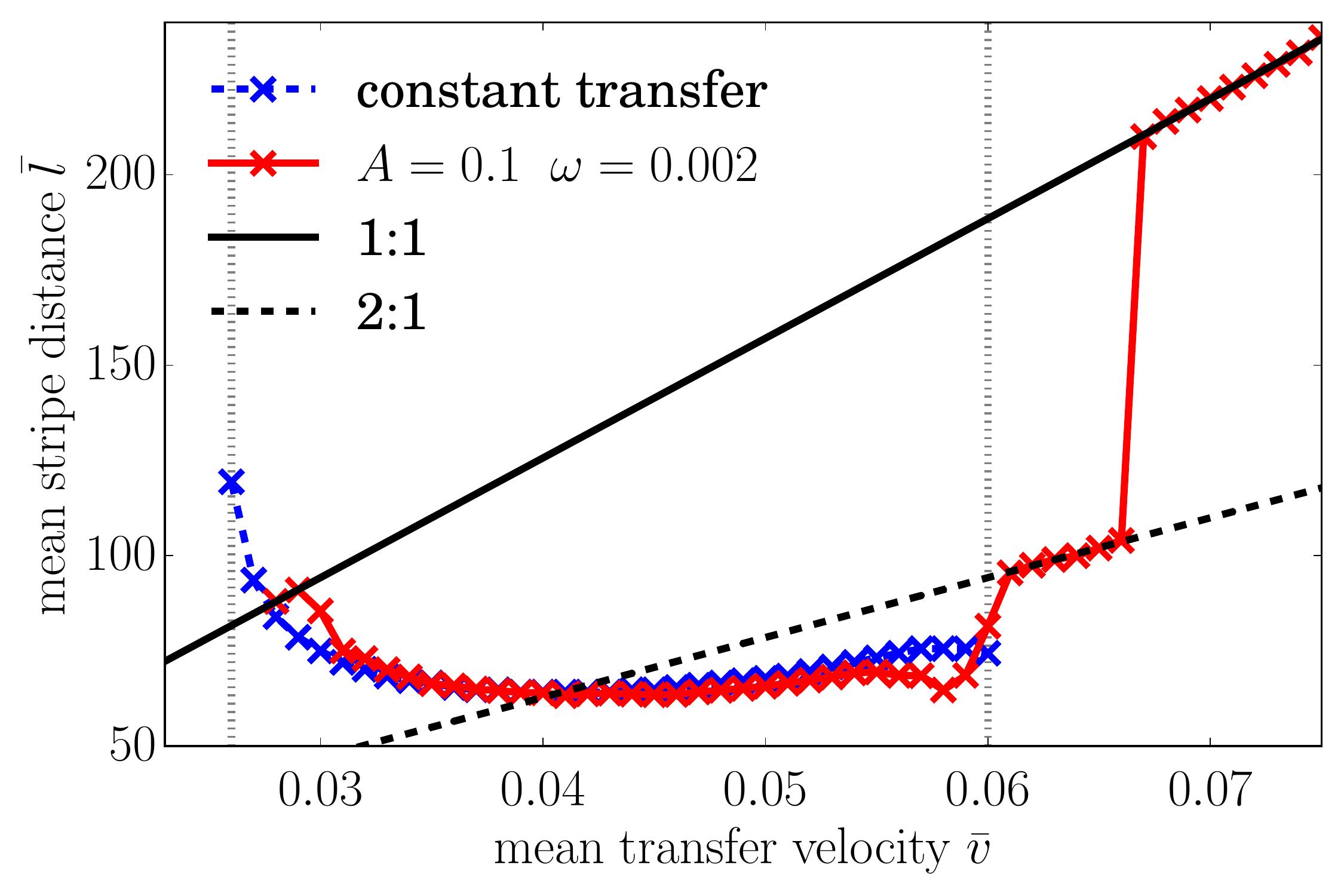}
	\caption{\label{fig:1d_v_periodic_l_synchrolines} Shown is the dependence of mean stripe distance $\bar{l}$ on mean transfer velocity $\bar{v}$ for horizontal stripe patterns in the 1D case.  Time-periodic forcing, i.e., Eqs.~\eqref{eq:full_ch_eq}-\eqref{eq:ic} with forcing frequency $\omega = 0.002$ and amplitude is $A = 0.1$ (red crosses) is compared to the case of constant transfer velocity $v = \bar{v}$ (blue crosses). The linear behavior of $\bar{l}$ [Eq.~\eqref{eq:sync}] in the case of complete synchronization is shown for $W = 1$ (solid black line) and $W = 2$ (dashed black line). Dotted vertical lines denote the limits of the natural patterning range. }
\end{figure}
Using results from the unforced system and Eq.~\eqref{eq:sync} we estimate appropriate parameter values for the forcing frequency $\omega$ and then choose a forcing amplitude $A$ depending on how large the influence of the forcing term shall be. Fig.~\ref{fig:1d_v_periodic_l_synchrolines} compares dependencies of mean stripe distance $\bar{l}$ on mean transfer velocity $\bar{v}$ for the system with time-periodic forcing and the system with constant transfer velocity. The straight black lines correspond to the linear behaviour of $\bar{l}$ predicted by Eq.~\eqref{eq:sync} for complete $1:1$ (solid line) and $2:1$ (dashed line) synchronization. Therefore, overlapping ranges between black and red lines represent extensive regions of frequency, i.e., stripe distance entrainment. Note that the entrainment does not exclusively happen in these regions. In fact, each depicted solution is entrained to the external forcing but with a $W$ which is composed of larger integers because the detuning is too large for low $W$. Apart from this, within the patterning regime of the unforced system $\bar{l}$ follows $l_{\mathrm{nat}}$ (blue crosses) very closely. This is not surprising since it is the natural stripe distance. But when the detuning is small enough the forcing can push $\bar{l}$ away from its natural value onto the value imposed by synchronization as can be seen from the deviation of blue and red lines. Therefore, there exists a competition between $l_{\mathrm{nat}}$ and synchronized $\bar{l}$. This becomes clear in the following observation as well: The periodic forcing extends the patterning regime towards larger and smaller $\bar{v}$ as compared to the one in the natural case, indicated in Fig.~\ref{fig:1d_v_periodic_l_synchrolines} by vertical dotted lines. 
In the natural case without forcing, beyond the end of the patterning regime (marked by vertical dotted lines), homogeneous deposition occurs. Thus, there is no competing natural stripe distance $l_{\mathrm{nat}}$ and in the forced case low ordered $W$ prevail in the form of extensive synchronization regions. 
\begin{figure}[tb]
	\includegraphics[width = 0.49\textwidth]{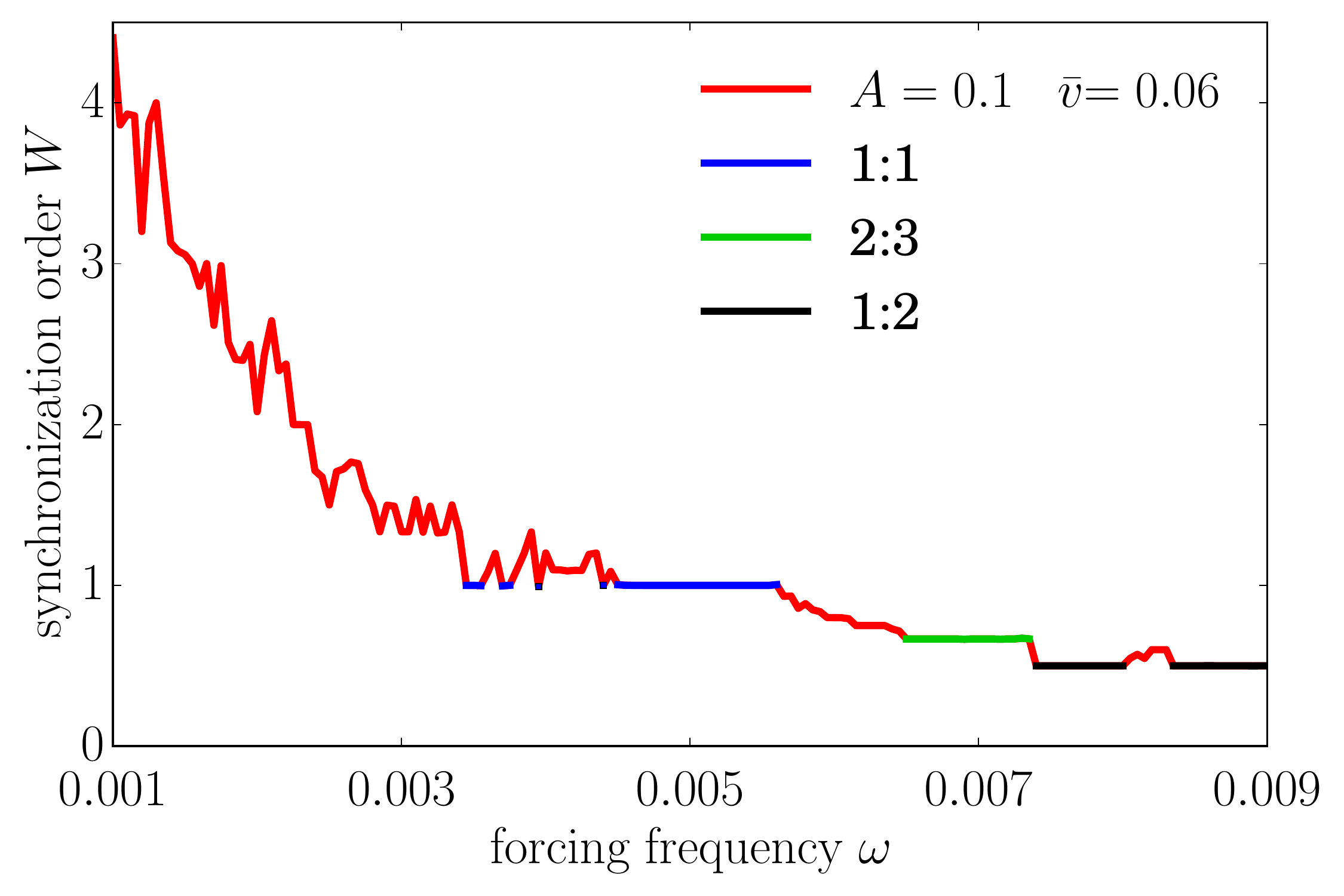}
	\caption{\label{fig:1d_w_periodic_W} The dependence of the synchronization order $W$ on the forcing frequency $\omega$ for horizontal stripe patterns in the 1D case with $A = 0.1$ and $\bar{v} = 0.06$ shows a typical devil's staircase. Plateaus for 1:1 (blue), 2:3 (green) and 1:2 (black) synchronization are particularly highlighted.}
\end{figure}

\begin{figure}[tb]
	\includegraphics[width = 0.49\textwidth]{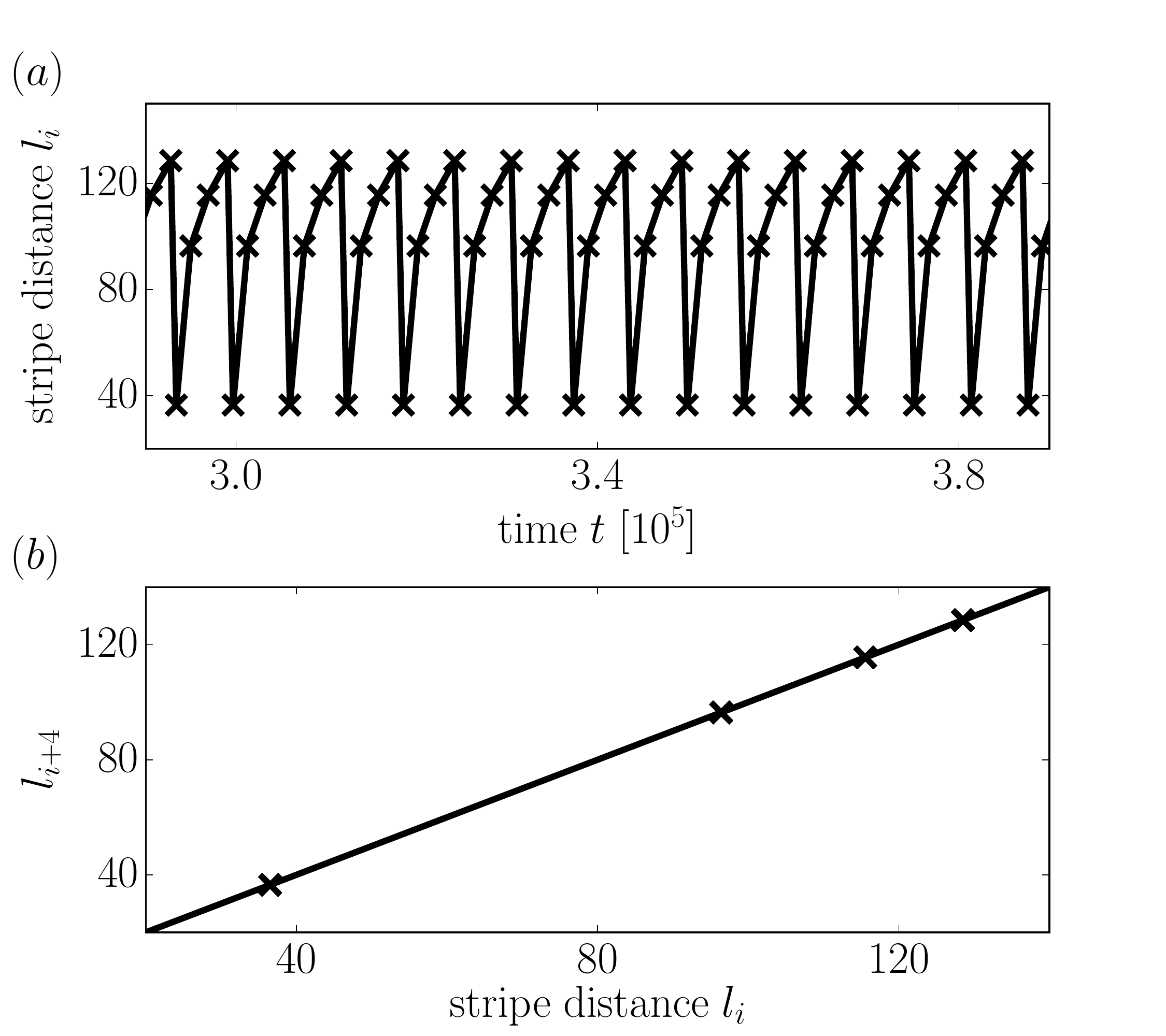}
	\caption{\label{fig:1d_v_periodic_return} (a) Time evolution of the stripe distance $l_i$ for horizontal stripe patterns in the 1D case with $A=0.1$, $\omega = 0.003$ and $\bar{v} = 0.06$ (b) and its fourth return map. The solution exhibits prime periodicity $4$ and $W = 4:3$. }
\end{figure}

We have seen that at the chosen $\omega$ and $A$, in the natural patterning range the dependence on $\bar{v}$ strongly correlates with the behavior of the natural (i.e., unforced) system. Next we consider $\omega$ and $A$ as control parameters. In Fig.~\ref{fig:1d_w_periodic_W} the synchronization order $W$ is given as a function of $\omega$ at fixed $A = 0.1$ and $\bar{v} = 0.06$. Overall, $W$ decreases with increasing $\omega$ which is not surprising since \eqref{eq:sync} shows that $W$ depends inversely on $\omega$.
In this representation, plateaus of nearly constant $W$ correspond to regions of synchronization. The larger plateaus usually result from low order synchronization like $1:1$ (blue), $2:3$ (green) or $1:2$ (black) resonnances. However, normally the integers that compose $W$ are large and the widths of the corresponding plateaus are very small. Therefore, as before, the system synchronizes at each parameter value. This behavior is very common for synchronized systems and known as ``devil's staircase'' \cite{Mandelbrot:1977}. The reader is referred to Fig.~\ref{fig:1d_v_periodic_return} below for a visualization of the synchronization behavior away from plateaus. Details regarding the employed methods are mentioned later on when discussing individual stripe distances. 
\begin{figure}
	\includegraphics[width = 0.5\textwidth]{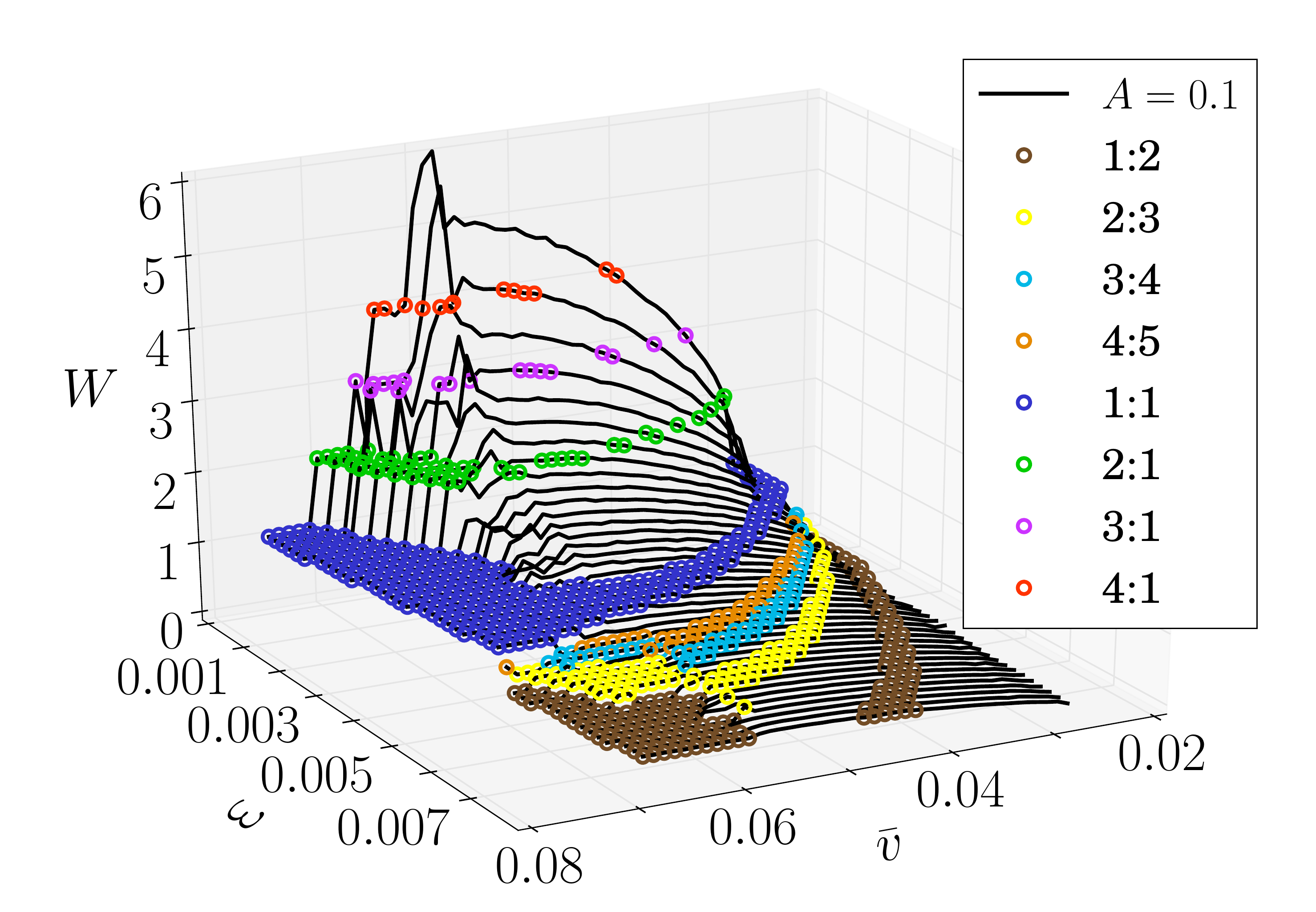}
	\hbox
	{
	\hspace{-2cm}
	\includegraphics[width = 0.6\textwidth]{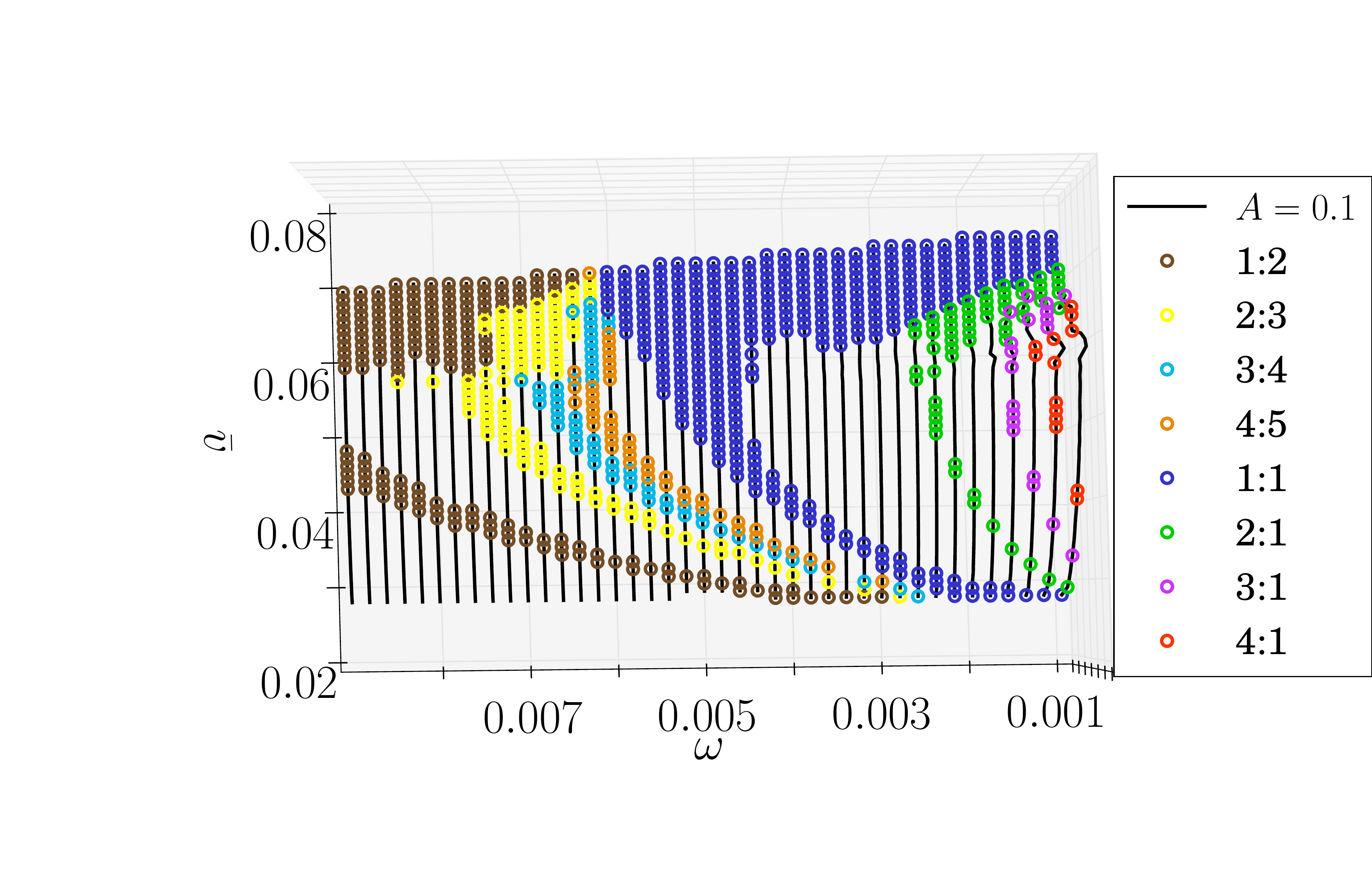}
	}
	\setlength{\abovecaptionskip}{-20pt}
	\caption{\label{fig:1d_vw_periodic_a01_W_001} Shown is the dependence of the synchronization order $W$ on the mean transfer velocity $\bar{v}$ and the forcing frequency $\omega$ at fixed $A = 0.1$ for horizontal stripe patterns in the 1D case. Top and bottom panel provide different views of the same data. Several plateaus at low order $W$ ranging from 1:2 to 4:1 are distinctively colored. The full patterning range in terms of $\bar{v}$ is covered, i.e., beyond the ends of the black lines, homogeneous transfer occurs. A video showing the plot from different perspectives is provided in the ancillary files. }
\end{figure}

To obtain an overview of the system behaviour across the parameter space we perform an extensive parameter scan in $\bar{v}$ and $\omega$. The result is given in Fig.~\ref{fig:1d_vw_periodic_a01_W_001}. Black lines represent occurring stripe patterns whereas coloring emphasizes the extensive synchronization regions for $W = 1:2$ to $4:1$. When following the data either at constant $\omega$ or at constant $\bar{v}$, the respective features of Figs.~\ref{fig:1d_v_periodic_l_synchrolines} and \ref{fig:1d_w_periodic_W} are recovered. Naturally, the periodic solutions still exist outside of the shown $\omega$-ranges. The step-like behavior close to the edges, e.g., at the large-$\bar{v}$ border of the 1:1 region, is an artifact resulting from the discretization of the control parameters. The previously mentioned extension of the patterning regime towards small velocities also becomes evident. This extension is larger for smaller $\omega$. It is remarkable that the synchronization regions of low ordered $W$ cover quite large parts of the investigated parameter plane, cf.~Fig.~\ref{fig:1d_vw_periodic_a01_W_001}~(bottom). 
This allows for some progress in predicting emerging stripe distances in real experiments, i.e., once the synchronization order of a pattern has been determined, the stripe distance for nearby parameters can be predicted using \eqref{eq:sync}.

\begin{figure}[tb]
	\includegraphics[width = 0.49\textwidth]{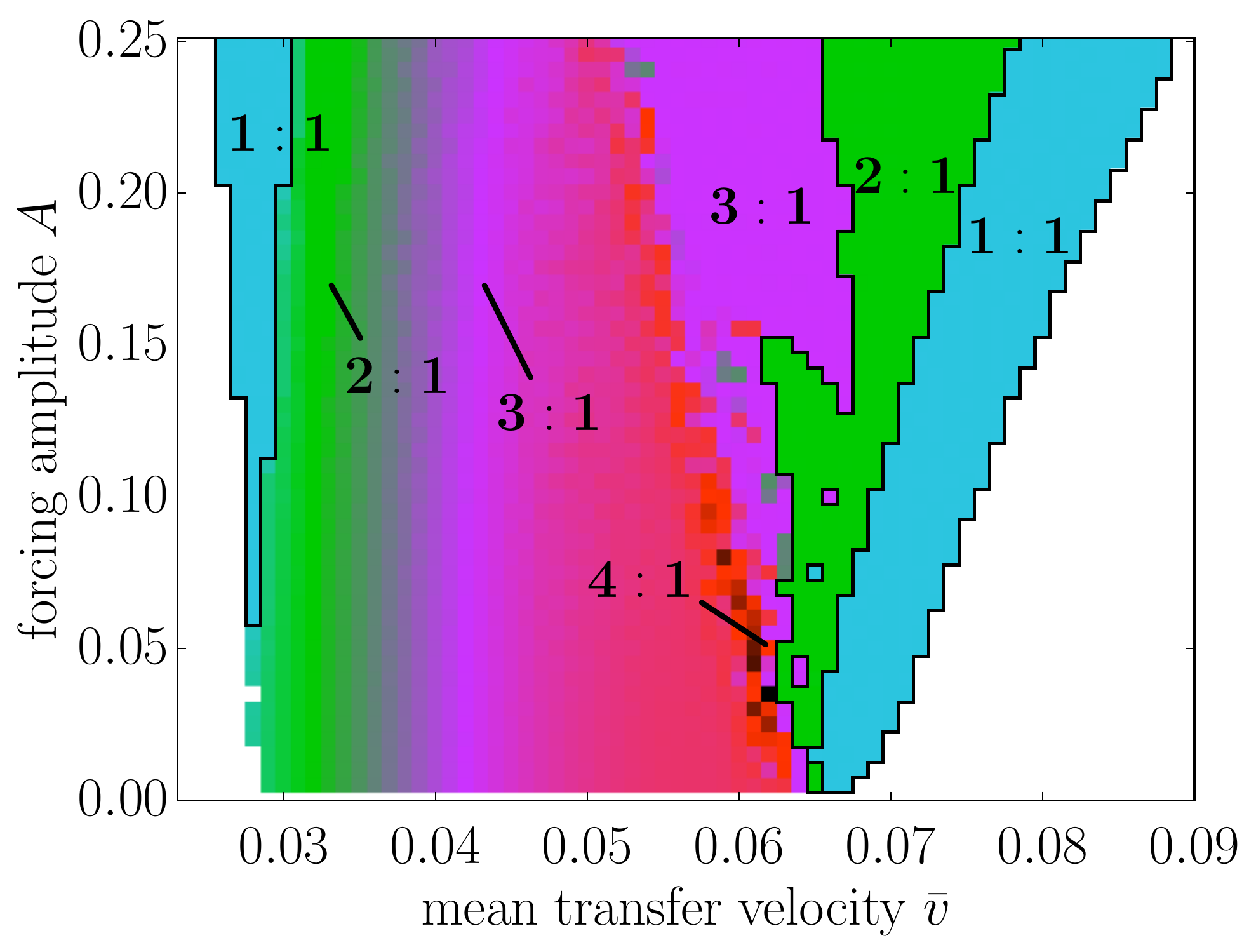}
	\caption{\label{fig:1d_va_arnold} Shown are Arnold tongues in the dependence of the synchronization order $W$ on the mean transfer velocity $\bar{v}$ and the forcing amplitude $A$ at fixed $\omega = 0.0014$ for horizontal stripe patterns in the 1D case. Each point represents a time-periodic solution. The synchronization order $W$ is indicated by the color scheme that is identical to the one in Fig.~\ref{fig:1d_vw_periodic_a01_W_001} for $W>1$. Here, $W<1$ does not occur.}
\end{figure}

We have shown that both, $\bar{v}$ and $\omega$, can be used to switch between different synchronization orders $W$. If instead one changes the forcing amplitude $A$, a further effect is observed, as illustrated in Fig.~\ref{fig:1d_va_arnold} that presents the dependence of $W$ on $A$ and $\bar{v}$ at fixed $\omega = 0.0014$. The color coding of the synchronization plateaus is as in Fig.~\ref{fig:1d_vw_periodic_a01_W_001} for $W > 1$. Analogeous calculations for the $(\omega, A)$ parameter plane yield similar results. As expected, for $A\approx 0$, the synchronization regions are small because $\bar{l}$ closely follows $l_{\mathrm{nat}}$, causing the impression that $W$ changes continuously. At larger $A$ we discern large regions of rather uniform $W$, in particular, for synchronisation of low order, e.g., $n:1$, $n\leq 4$. These regions form the well-known Arnold tongues \cite{Arnold:2009}. Most of the tongues emerge close to the borders of the patterning regime in the case without forcing ($\bar{v} \in \{0.026,0.06\}$ for $A = 0$) and expand their $\bar{v}$-range with increasing $A$. The transitions between tongues of different $W$ are particularly sharp not too far from these borders as was already observed in Fig.~\ref{fig:1d_v_periodic_l_synchrolines}. Starting at $\bar{v} = 0.06$ and $A = 0$, a series of abrupt changes in $W$ with intermittent small regions with $4:1$ synchronization marks the border between low-ordered entrainment to the forcing and entrainment to the natural stripe distance. Note, that a very strong increase of $A$ will eventually lead to the system being perpetually synchronized $1:1$~(not shown). 

\begin{figure}[tb]
\includegraphics[width = 0.49\textwidth]{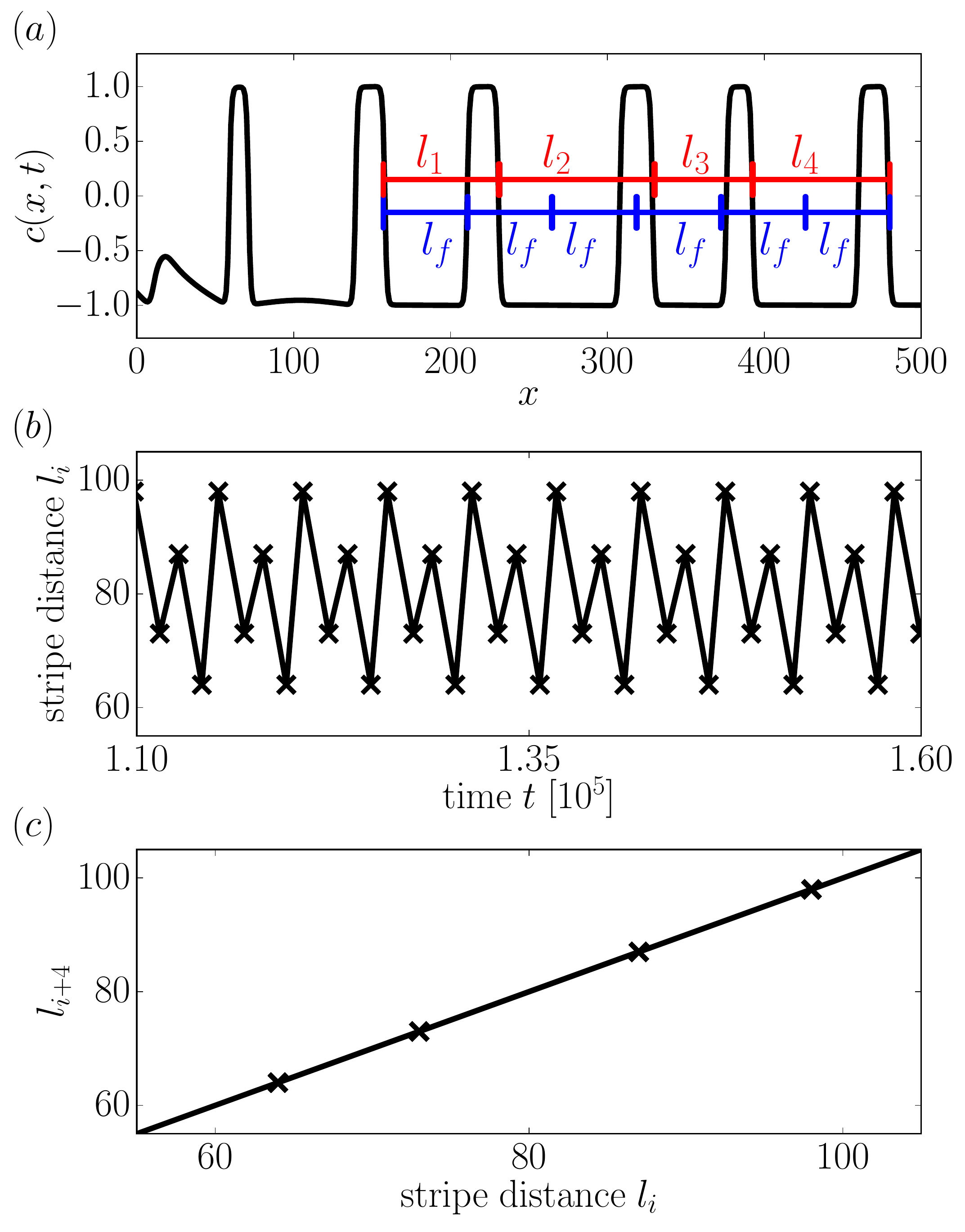}
\caption{\label{fig:1d_2_3} A $2:3$ synchronized horizontal stripe patterns in the 1D case as solutions of \eqref{eq:full_ch_eq}-\eqref{eq:ic} with $A = 0.1$, $\omega = 0.007$ and $\bar{v} = 0.06$. (a) A snapshot of the solution is presented. The $l_i$ (red) denote the four individually occurring stripe distances and $l_f$ is the forced stripe distance for $1:1$ synchronization (cf. \eqref{eq:sync}). (b) Time evolution of $l_i$. (c) Fourth return map of the individual stripe distances $l_i$, indicating that the solution is $4$-periodic.}
\end{figure}

Up to now, we have only reported on mean stripe distance $\bar{l}$ (see Fig.~\ref{fig:1d_v_periodic_l_synchrolines}), but have abstained from statements about the individual stripe distances $l_i$ with $i\in \mathbb{N}$. For the investigated parameter sets we have not observed solutions consisting of quasiperiodic or chaotic sequences of $l_i$. Fig.~\ref{fig:1d_2_3} (a) shows a snapshot of a particular 2:3 synchronized solution where one discerns four different stripe distances $l_i$ which periodically repeat. Fig.~\ref{fig:1d_2_3}~(b) gives 
the sequence of the corresponding stripe distances $l_i$ over time measured at $x = \frac{4}{5} L$ as the developed stripe pattern passes. Using this information, in Fig.~\ref{fig:1d_2_3}~(c) we construct the fourth return map $l_{i+4}(l_{i})$. Only four different points are visible confirming that the solution is indeed $4$-periodic. We show another return map in Fig.~\ref{fig:1d_v_periodic_return} for forcing frequency $\omega = 0.003$ which is outside of the large plateaus visible in Fig.~\ref{fig:1d_w_periodic_W}. This type of behavior is found for most synchronization orders. However, for those with large integer values the overall periodicity can become quite large of the order of $10^1-10^2$. In these cases the periodicity is not calculated as it requests too many computing resources. Here, only the periodicity of $1:n$ synchronization is numerically determined, i.e., cases where all stripe distances are identical $l_i = l$, though there are cases where the solutions with $W = 1$ do exhibit $l_i, i \in \{1,2\}$. 

\begin{figure}[tb]
	\includegraphics[width = 0.49\textwidth]{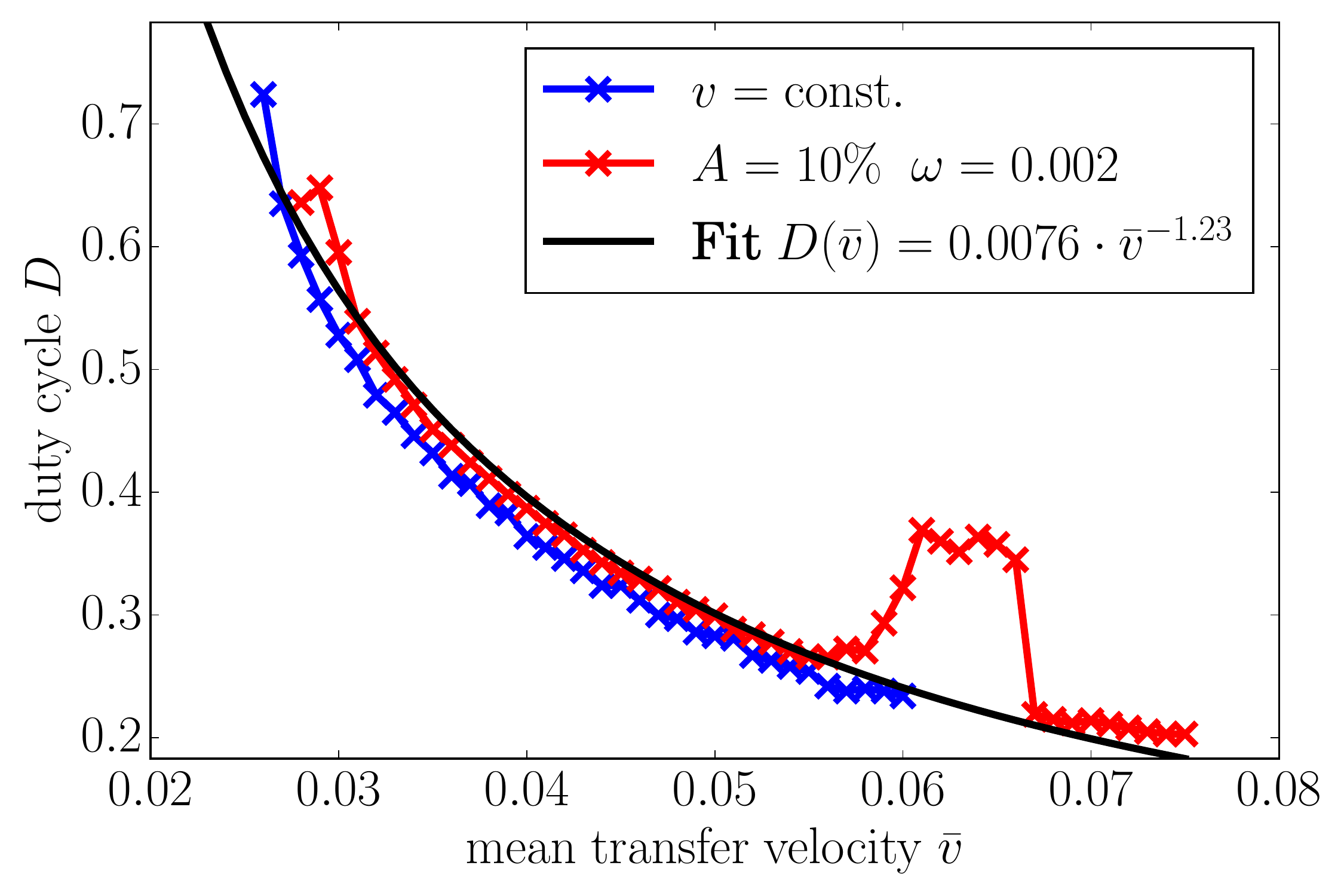}
	\caption{\label{fig:1d_v_periodic_duty} Shown is the dependence of the average duty cycle $D$ on the mean transfer velocity $\bar{v}$ for horizontal stripe patterns in the 1D case in the forced case with $A = 0.1$ and $\omega = 0.002$ (red) in comparison to the unforced case (blue). }
\end{figure}

The relative distribution of LC and LE phases is quantified employing the average duty cycle $D$ defined in Sec.~\ref{sec:solmeas}. The dependence of $D$ on $\bar{v}$ is shown in Fig.~\ref{fig:1d_v_periodic_duty} for $A = 0.1$ and $\omega = 0.002$. There, we see that similarly to the unforced system, the forced system shows inverse behavior. The main difference is the extension of the patterning regime and the occurrence of frequency locking at higher velocities. The inverse behavior can be understood by recollecting that with increasing $\bar{v}$ the system moves closer and closer to homogeneous LE phase deposition. Therefore, considering the definition $D = \overline{l_{\mathrm{LC}}/l}$, the part with LE phase increases and the part with LC phase decreases with increasing $\bar{v}$. In Fig.~\ref{fig:1d_v_periodic_duty} a power law with negative exponent is fitted to the data of the forced system for the purpose of comparison. 
The jumps visible at high velocities result from switches in synchronization order. 

\section{Effects of temporal forcing in two dimensions}\label{sec:2d}
\begin{figure}[tb]
	\includegraphics[width = 0.49\textwidth]{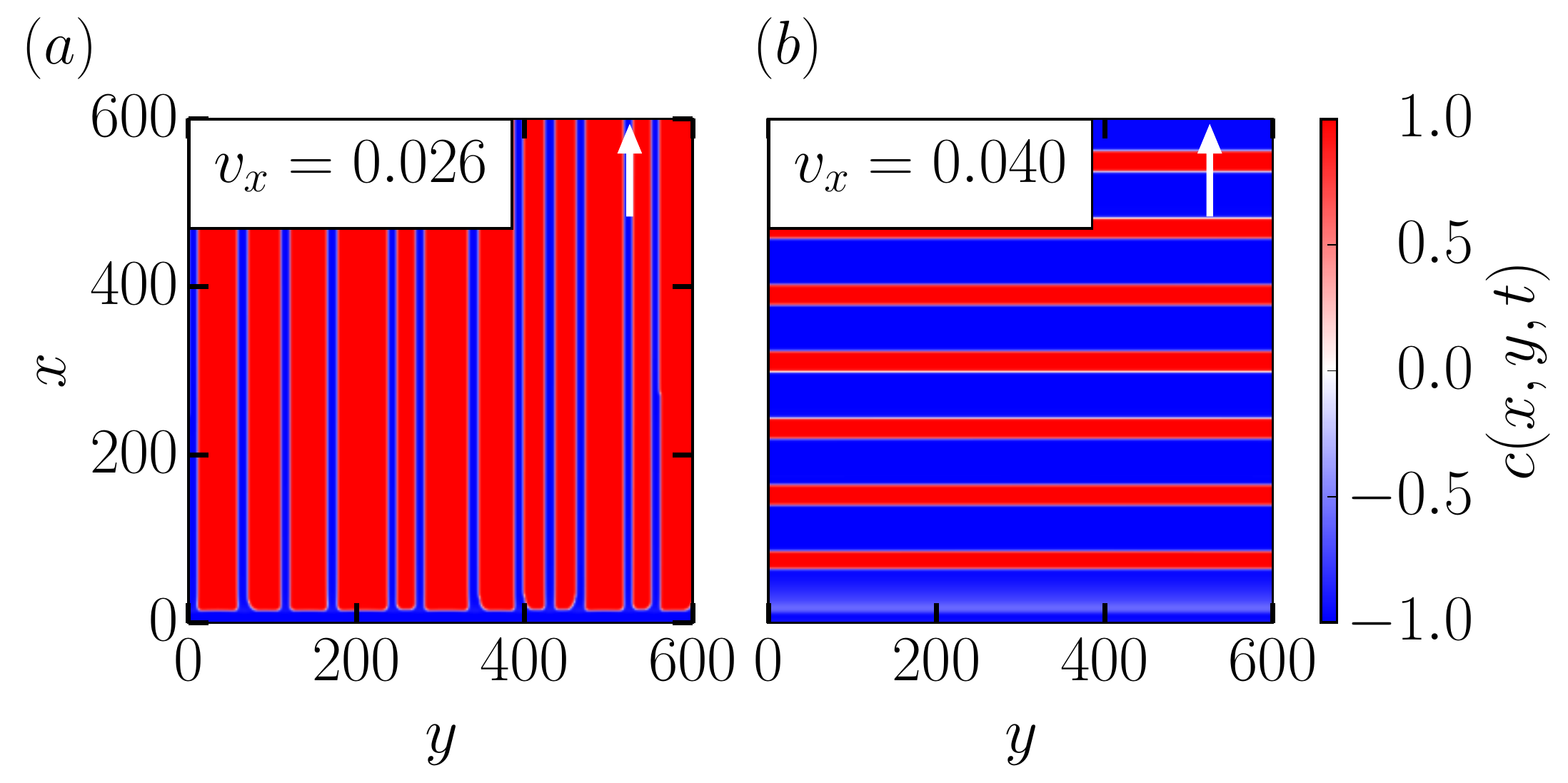}
	\caption{\label{fig:2d_vconst_soltypes} Examples of two-dimensional solutions of \eqref{eq:full_ch_eq} in the case of constant transfer velocity ($A = 0$). The transfer direction is denoted by the respective white arrow. (a) Vertical stripes arising at $\bar{v} = 0.026$ after an initial transient has died out. (b) Horizontal stripes at $\bar{v} = 0.04$. }
\end{figure}

Next, we report on additional effects found when applying time-periodic forcing in the 2D system. First, we recapitulate results known from previous work on deposition \cite{KoepfGurevichFriedrich:2011,WilczekGurevich:2014}. 
At small velocities there exists a small region of transversal instability which quickly destabilizes emerging horizontal stripes that then evolve into vertical stripes (Fig.~\ref{fig:2d_vconst_soltypes}~(a)).
In this section the transfer direction $x$ is vertical with the plate moving towards the top and the horizontal coordinate is denoted by $y$. 
Furthermore, Ref.~\cite{WilczekGurevich:2014} includes spatial forcing in the form of spatial modulations of the SMC-term (cf.~\eqref{eq:zeta}) in the frame of the moving plate, that represent prestructures on the substrate. As a result, the transversal instability can be suppressed or amplified depending on the orientation of the modulation. If the prestructure stripes are oriented vertically, also oblique LE/LC stripes and lattice structures can form in addition to the vertical stripes. In the case of horizontally oriented prestructure stripes, horizontal stripes remain the only emerging structure.

\begin{figure*}[tb]
	\includegraphics[width = \textwidth]{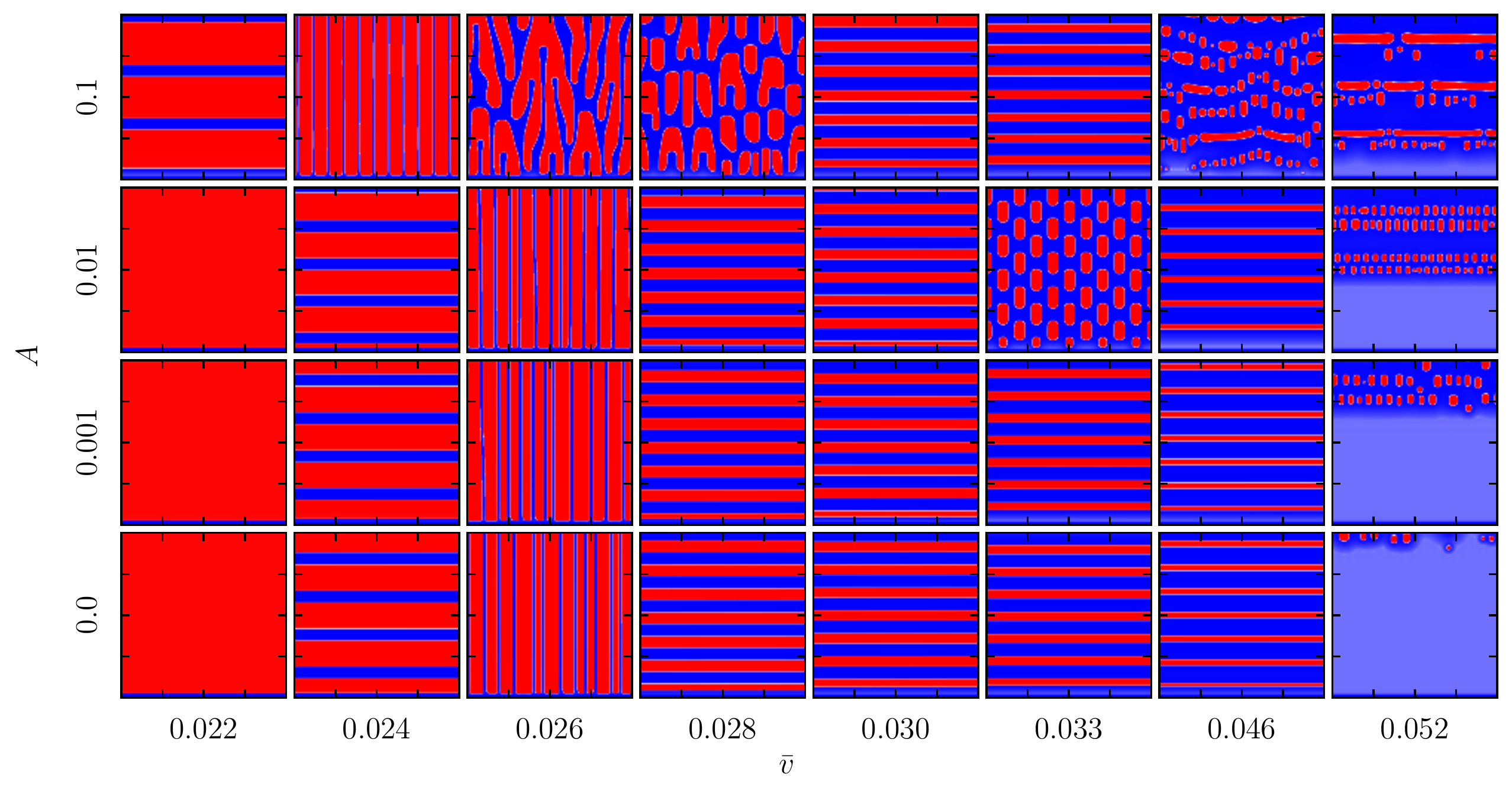}
	\caption{\label{fig:2d_va_noise} Examples of snapshots of 2D solutions of \eqref{eq:full_ch_eq}-\eqref{eq:ic} in the plane spanned by mean transfer velocity $\bar{v}$ (left to right) and forcing amplitude $A$ (bottom to top) at $\omega = 0.002$. Shown are the states at end of simulation, namely at time $t_\mathrm{end}= 10\tau$, where $\tau$ is the time needed for the substrate to be advected by one domain size $L$. This limit is increased for transversally inhomogeneous patterns. The initial condition is a simple meniscus  with small amplitude white noise as in \eqref{eq:ic}. The transfer direction is from bottom to top. Note that to capture the most interesting solutions, the shown $\bar{v}$ intervals are not equidistant. }
\end{figure*}

From this, one could expect that transversally inhomogeneous structures do not emerge with our time-periodic forcing term \eqref{eq:forcing} since its application is homogeneous with respect to the $y$-direction and the plate velocity is always strictly pointing in the $x$-direction. This should promote the formation of horizontal stripes and suppress the transversal instability as in the case of horizontal prestructure stripes. Interestingly, this is not the case. Using initial conditions as in the 1D case that are homogeneously extrapolated into the $y$-direction, when a small amount of white noise is initially added the emerging stripes still exhibit a transversal instability in a certain range of plate velocities. Without initial noise, the results of the 1D case are still valid, i.e., horizontal stripes form under corresponding synchronization effects. For the case with initial noise, Fig.~\ref{fig:2d_va_noise} shows snapshots of the converged solutions of \eqref{eq:full_ch_eq}-\eqref{eq:ic} for different amplitudes $A$ and velocities $\bar{v}$ at $\omega = 0.002$. At small $A = 0.001$, the system behaves very similar to the unforced case.  It displays a small region of transversal instability at small $\bar{v}\approx 0.026$ where ultimately vertical stripes form. In contrast to the case of forcing by prestructuring \cite{WilczekGurevich:2014}, here the transversal instability is not suppressed. With increasing $A$, the instability is amplified as indicated by the emergence of further transversally inhomogeneous patterns like branched and lattice structures. This is surprising since one would not expect the \emph{temporal} forcing to affect the transversal direction. Upon approaching the upper border of the patterning range, e.g.,  at $A = 0.01$ for $\bar{v}\approx 0.046$, the horizontal stripes break up while retaining some periodicity due to the system's natural stripe distance and/or the forcing. The discontinuous occurrence of these different patterns in parameter space leads us to the conjecture that the initial condition is very important, e.g., to obtain transfer patterns that are not horizontal stripes it needs to be sufficiently transversally inhomogeneous.

\begin{figure}[tb]
	\includegraphics[width = 0.49\textwidth]{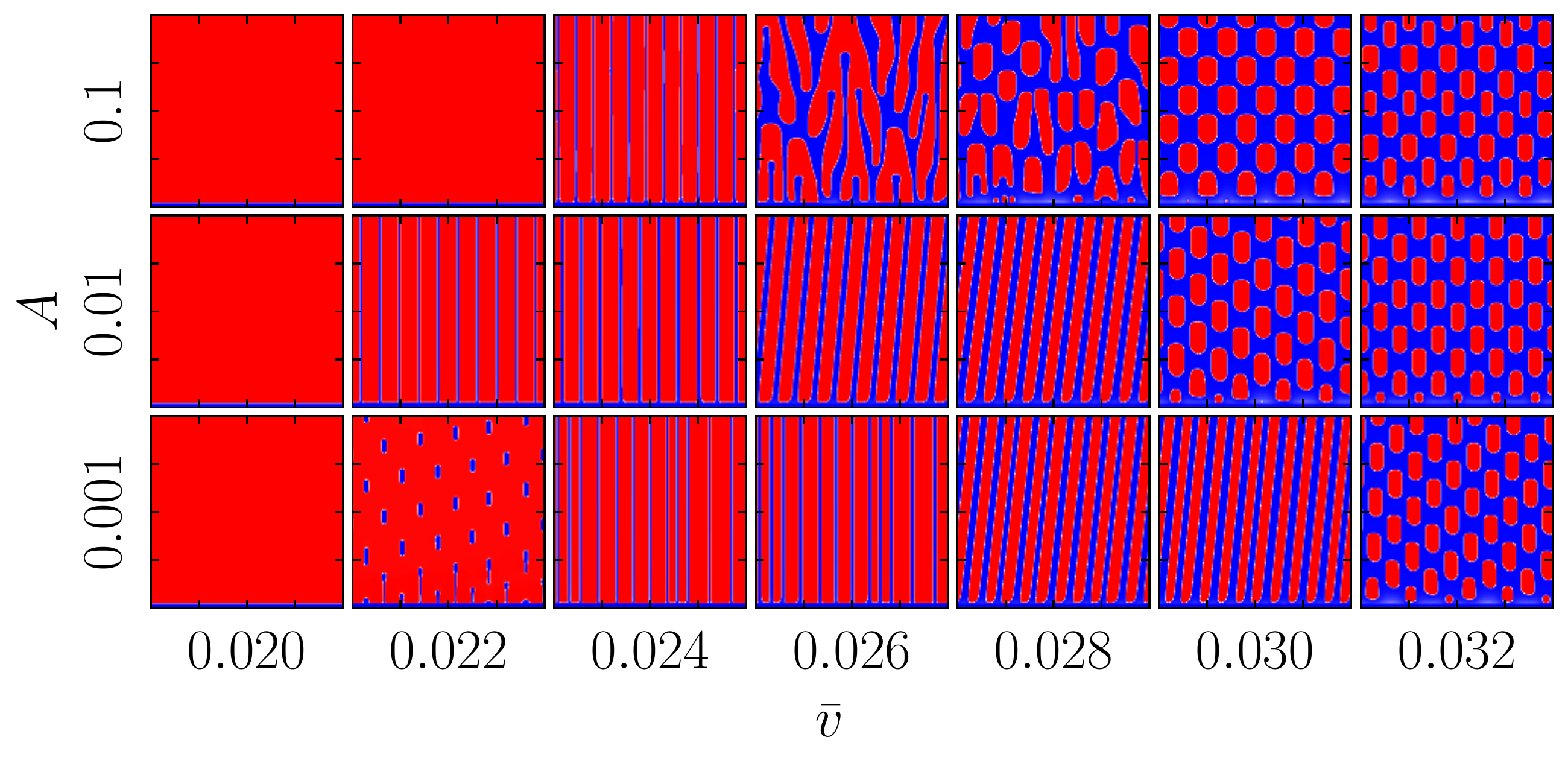}
	\includegraphics[width = 0.49\textwidth]{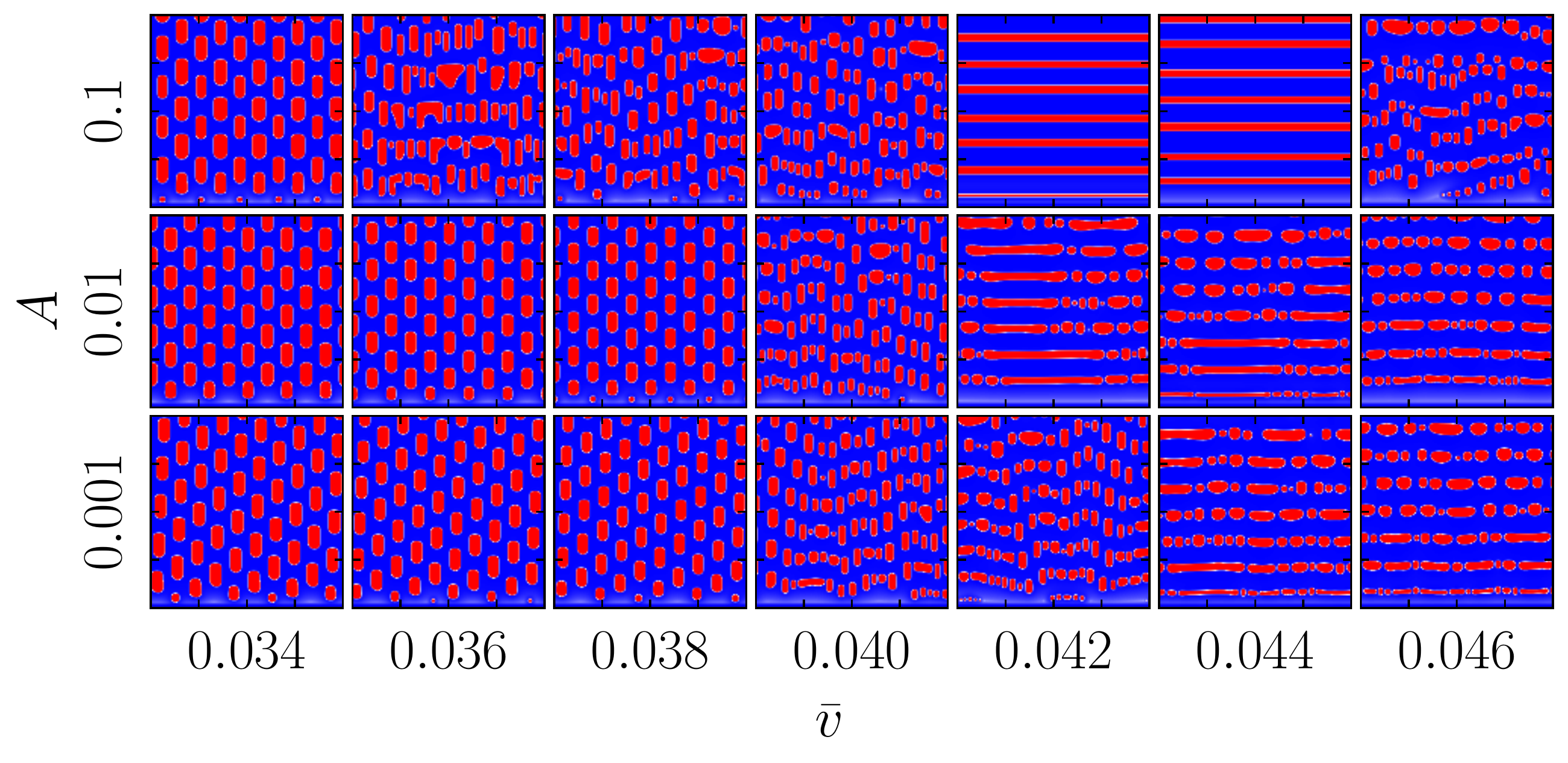}
	\caption{\label{fig:2d_va_transinsta} Array of snapshots of 2D solutions of \eqref{eq:full_ch_eq}-\eqref{eq:ic} in the plane spanned by mean transfer velocity $\bar{v}$ (left to right) and forcing amplitude $A$ (bottom to top) at $\omega = 0.002$. Shown are the states at end of simulation, namely at time $t_\mathrm{end}= 10\tau$, where $\tau$ is the time needed for the substrate to be advected by one domain size $L$. This limit is increased for transversally inhomogeneous patterns. For each run, the final solution of the previous run is used as initial condition whereby the very first initial condition is a vertically striped pattern. The transfer direction is from bottom to top. Here, the shown $\bar{v}$ intervals are equidistant }
\end{figure}

To investigate the extent of the occurrance of transversally inhomogeneous structures, a procedure of simple numerical continuation was performed: for each control parameter set the final solution of the previous set is used as initial condition. The very first set is $\bar{v} = 0.024$ from where the velocity is increased or decreased. The resulting patterns are shown in Fig.~\ref{fig:2d_va_transinsta}, and confirm the above conjecture. Depending on the extent of the spatial inhomogeneity in the $y-$direction, vertical, oblique, lattice structures or broken horizontal stripes may emerge. Interestingly, with this continuation scheme the pattern types become increasingly complex with increasing $\bar{v}$,  i.e., two fourier modes (i.e., wavevectors) become unstable, resulting in the persistent emergence of lattice structures in a large part of the parameter plane, e.g., see at $A = 0.01$ the transition from $\bar{v} = 0.028$ to $\bar{v} =  0.030$. 
This shows that the employed forcing does not only change the overall patterning range but the parameter ranges where the different pattern types occur. In the case of oblique stripes it is one mode with both wavenumber components being nonzero that becomes unstable. Still, at $A = 0.1$, $\bar{v} = 0.044$ the large forcing suppresses the transversal instability, forming horizontal stripes again. This suggests that the forcing amplitude may neither be too low nor too high if transversally inhomogeneous patterns are desired. During these calculations hysteresis phenomena have been found as well, indicating multistability between different solutions (this is not further analyzed here). \\
Usually, any defects visible in the snapshots shown in Fig.~\ref{fig:2d_va_noise} indicate patterns which are not yet fully developed, e.g., the two leftmost stripes of $A = \{0.001,\,0.01\}, \,\overline{v} = 0.026$, which are not entirely parallel yet, or the part above the homogeneous section of $A = 0.001,\, \overline{v} = 0.052$. In particular, transversally inhomogeneous patterns have very long transients because the developing periodicity of the patterns and the domain size are not commensurable. Hence, we limit the simulation time to $t_\mathrm{end}= 10\tau$ where $\tau$ is the time needed for the substrate to be advected by one domain size $L$, making parameter scans reasonably fast. This limit is set larger for transversally inhomogeneous patterns. However, it is also possible that patterns are fully developed in the vertical direction but not in the horizontal one, see e.g., $A = 0.01, \,\overline{v} = 0.052$ in Fig.~\ref{fig:2d_va_noise}. This cannot be seen in the corresponding $L^2$ norm (not shown here). In contrast, the vertical stripes at $A = 0.1,\,\overline{v} = 0.024$ are fully developed due to the strong forcing. 

\begin{figure}[h!]
	\includegraphics[width = 0.49\textwidth]{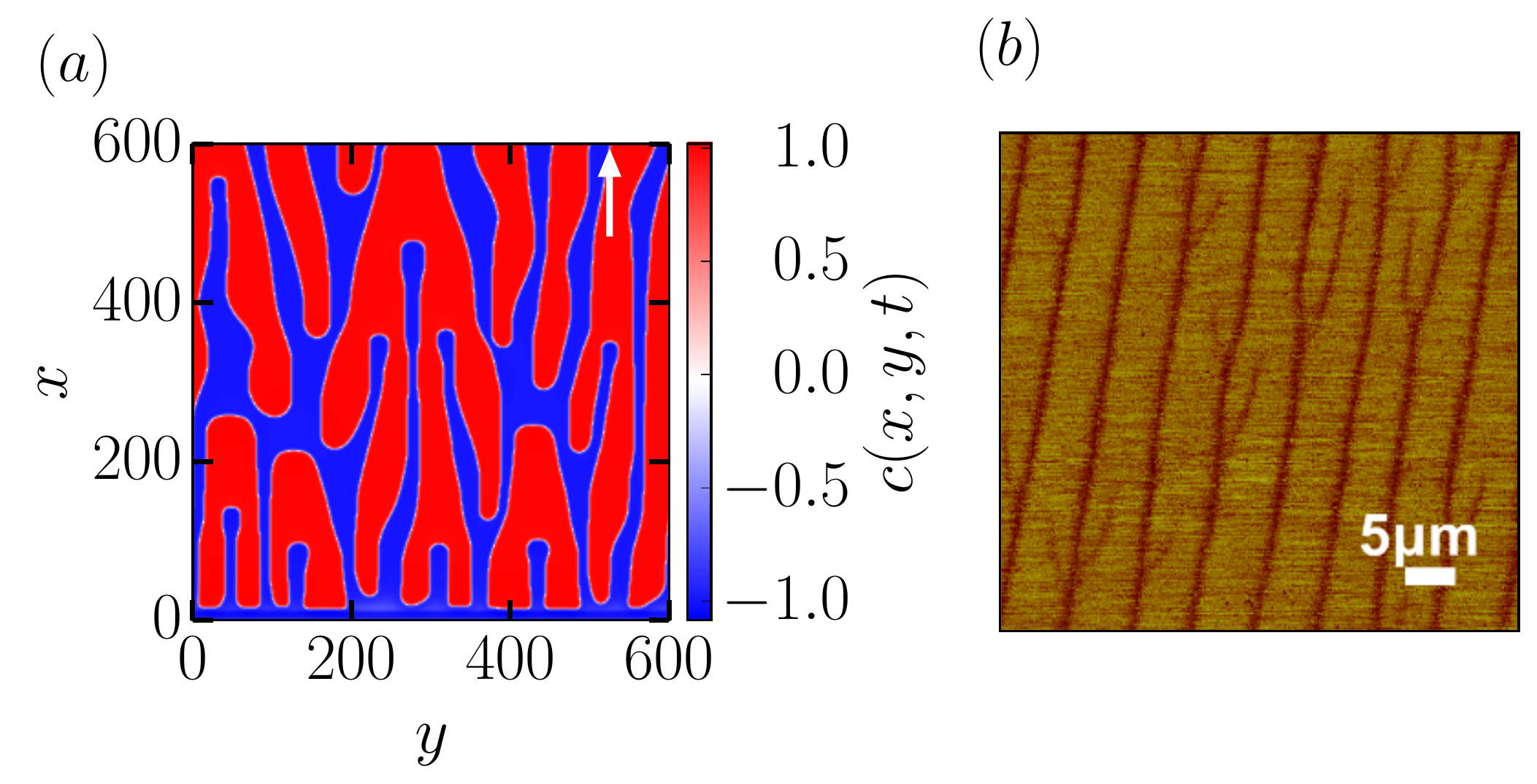}
	\caption{\label{fig:2d_transient}(a) Branched pattern in the case of time-periodic forcing with $A = 0.1$, $\omega = 0.002$, $\bar{v} = 0.026$. Blue and red domains represent the LE and LC phase, respectively. (b) AFM image of a branched pattern obtained for LB transfer at constant transfer velocity \cite{ZhuWilczekHirtz:2016}. Dark brown and yellow domains represent LE and LC phase, respectively. }
\end{figure}

Careful inspection of Fig.~\ref{fig:2d_va_transinsta} shows that another pattern type may occur consisting of a branched structure (cf.~$A = 0.1$, $\bar{v} = 0.026$). The panel is enlarged in Fig.~\ref{fig:2d_transient}~(a) and and paired with an experimental result (Fig.~\ref{fig:2d_transient}~(b)) measured at constant transfer velocity \cite{ZhuWilczekHirtz:2016}. The results resemble each other when identifying the red phase in the simulation with the yellow phase of the experimental image. However, from the time evolution of the $L^2$-norm of $c(x,y)$ (not shown here) it cannot be determined whether the simulated branched solution is asymptotically stable or whether it corresponds to a long-lived transient state. However, the similarity would then suggest that experimentally found branched structures may also be transients which would (for long enough plates) eventually become more regular patterns.

\begin{figure}[H]
	\includegraphics[width = 0.49\textwidth]{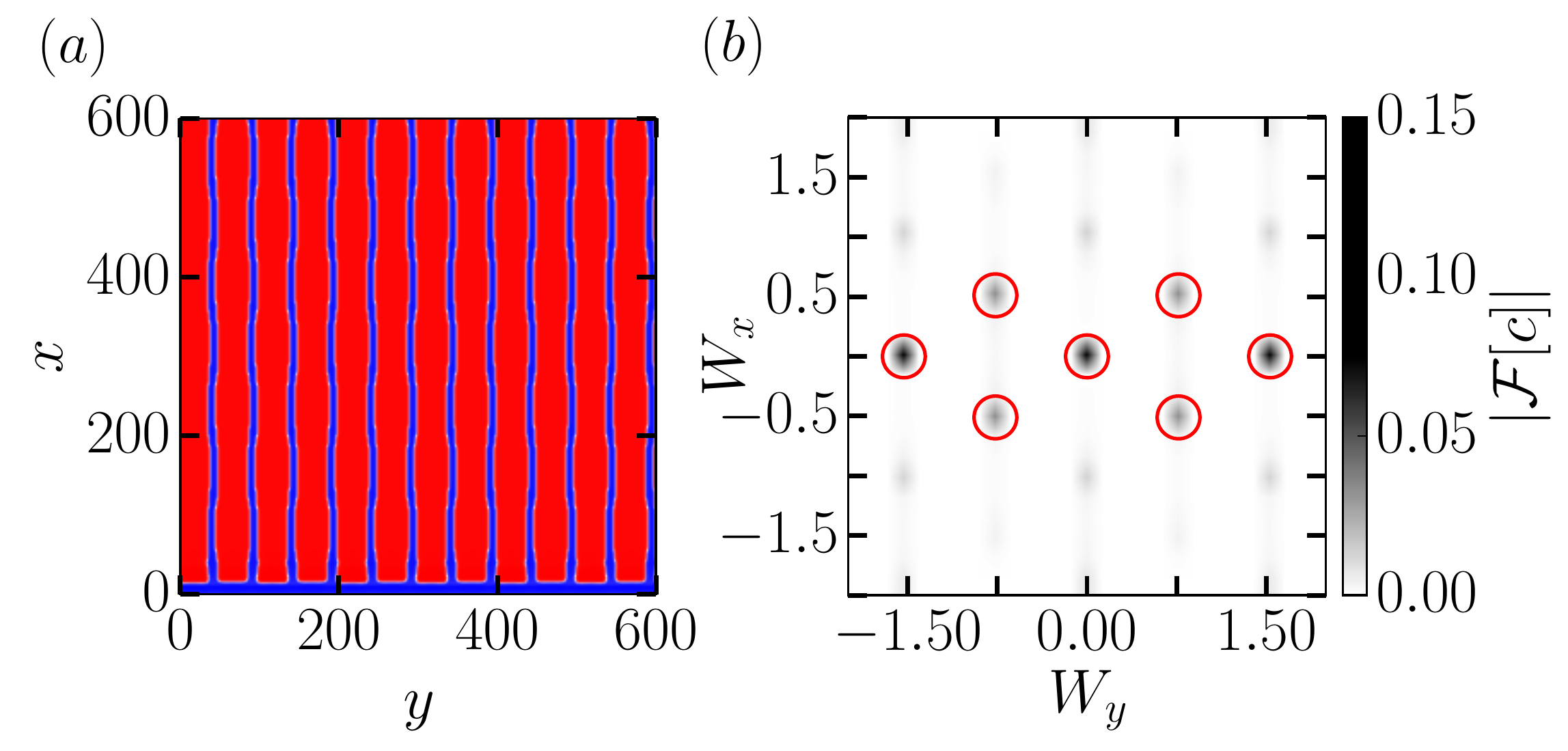}
	\caption{\label{fig:2d_pulsing}(a) Transversally modulated vertical stripes found in the case of time-periodic forcing with $A = 0.2$, $\omega = 0.002$, and $\bar{v} = 0.0244$. (b) Fourier transform $\mathcal{F}[c]$ of the pattern in the plane spanned by synchronization orders $W_y$ and $W_x$. The synchronization order can be directly related to a wavenumber according to Eq.~\eqref{eq:sync}. Each mode consists of one pair of opposite wavevectors, e.g., $(W_y, W_x) = (0.75, -0.5)$ and $(W_y, W_x) = (-0.75, 0.5)$ are one mode. The zero mode originates from the non-vanishing mean concentration of the pattern. }
\end{figure}

Within the region of vertical stripes it is also possible to find transversally modulated vertical stripes as shown in Fig.~\ref{fig:2d_pulsing}~(a). Such a modulation leads to a ``pulsing'' behaviour of the stripes where neighboring stripes are in anti-phase similar to interacting Plateau-Rayleigh instabilities of neighboring liquid ridges as seen in Sec. 4 of \cite{TBBB2003epje}. Here, this pattern shows frequency entrainment in both spatial directions which can be seen in the Fourier transform $\mathcal{F}[c]$ of the pattern given in Fig.~\ref{fig:2d_pulsing}~(b). There, the $k$-components in the $y$- and $x$-direction are given in the form of synchronization orders $W_y$ and $W_x$, respectively (using Eq.~\eqref{eq:sync}). The pattern consists of three superposed fourier modes where the two modes closest to the origin represent a rectangular pattern which is responsible for the pulsing while the $W_y = 1.5$ mode corresponds to the vertical stripes. The first two modes show $1:2$ synchronization in the $x-$ and $3:4$ synchronization in the $y$-direction, whereas the third mode exhibits $3:2$ synchronization in the $y$-direction only. Such an entrainment in the transversal direction is another indicator for the overly important influence of the forcing. 

\section{Conclusion and Outlook}\label{sec:conclusion}
A theoretical investigation of Langmuir-Blodgett transfer by means of a generalized Cahn-Hilliard model has been presented. We have shown that the inclusion of temporal forcing leads to a variety of synchronization effects for horizontal stripes, namely frequency entrainment, devil's staircase and Arnold tongues. By adjusting the forcing amplitude, frequency and mean transfer velocity it is possible to control the mean distance of deposited stripes, even managing synchronization orders larger than one, and also to increase the range of the patterning regime. There exists a competition between the system's natural stripe distance and the forced stripe distance. The former gains towards the center of the patterning  regime of the unforced system while the latter dominates towards its borders. Indeed, forcing extends the range of patterning. The results on synchronization effects and the range of patterning regime carry over from the one-dimensional to the two-dimensional case. Naturally, this is not the case for results related to the transversal instability that is an intrinsically 2D effect.

In 2D simulations it has been shown that the transversal instability which is present in the unforced system is not suppressed by the temporal forcing. Instead it is enhanced, contrary to expectations resulting from calculations with spatial forcing by means of prestructures~\cite{WilczekGurevich:2014}. The occuring pattern types include vertical and oblique stripes, branched structures, lattice structures and broken horizontal stripes. The parameter ranges where the various patterns occur can be changed by adjusting the forcing amplitude. Furthermore, transversally modulated vertical stripes were found to exist in a small parameter range. The particular shown example exhibits synchronization in both spatial directions, indicating two-dimensional synchronization similar to results obtained by spatial forcing~\cite{RuedigerPRL2003,manor2008wave,WilczekGurevich:2014}.

In order to gain a more thorough understanding of the system, the numerical analysis has to go beyond the direct numerical simulations presented here since it only provides stable solutions. In particular, the occurrence of multistability makes the detection of different solution branches a challenging task. Therefore, investigating this system by means of path-continuation and bifurcation theory is an important future step. First steps in using the path-continuation methods for two-dimensional partial differential equations for some related thin-film and Cahn-Hilliard-type systems were taken recently~\cite{Wilczek_M.:2017,Engelnkemper2019,Engeln16}, providing a basis for further research. 

\begin{acknowledgments}
We acknowledge partial support by Deutsche Forschungsgemeinschaft within PAK 943 (Project No. 332704749), and German-Israeli Foundation for Scientific Research and Development under Grant No. I-1361-401.10/2016. We thank Markus Wilczek for his preliminary work and very helpful discussions. 
\end{acknowledgments}
\FloatBarrier

\providecommand{\noopsort}[1]{}\providecommand{\singleletter}[1]{#1}%

\end{document}